\documentclass[12pt]{article}
\usepackage{xcolor}
\usepackage{amsmath}
\usepackage{authblk}
\usepackage{graphicx}
\usepackage{natbib}
\usepackage{amssymb}
\usepackage{caption}
\usepackage{subcaption}
\usepackage{tikz}
\usepackage[a4paper, total={6.5in, 8in}]{geometry}

\newcommand{\blue}[1]{#1}

\DeclareMathOperator*{\argmin}{arg\,min}

\begin{document}

\title{Enhancing Stellarator Accessibility through Port Size Optimization}

\author[1]{A. Baillod}
\author[1]{E. J. Paul}
\author[2]{T. Elder}
\author[1]{J. M. Halpern}
\affil[1]{Department of Applied Physics and Applied Mathematics, Columbia University, New York, New York 10027, USA}
\affil[2]{University of Maryland, College Park, Maryland 20737, USA}
\renewcommand\Affilfont{\itshape\small}
\maketitle

\begin{abstract}
Access to the plasma chamber in a stellarator reactor is essential for maintenance and diagnostics. However, the complex geometry of stellarator coils, often characterized by their strong twisting, can severely limit the space available for access ports. This study introduces a novel optimization approach in which access ports are represented as closed curves on the plasma boundary. By carefully selecting a set of objectives and penalties related to the access port, we demonstrate the first stellarator coil optimization explicitly targeting improved access port size. The trade-off between magnetic field quality and port size is analyzed through the Pareto front of their respective objectives. The optimal location of a port is explained using a current potential approach. Finally, we show that additional shaping coils, such as windowpane coils, can enable the crossing of the Pareto front to achieve superior configurations. 
\end{abstract}

\section{Introduction}
Stellarators are magnetic fusion devices that confine a plasma with a magnetic field generated by external coils. In a stellarator reactor, the extreme heat and neutron fluxes generated by the fusion process, progressively damage the plasma facing components and the reactor's blanket elements \citep{knaster_2016}. The strategy used to replace the damaged elements, called maintenance scheme (MS), is thus a crucial element of reactor design \citep{boozer_2021, boozer_2024}.

The MS has to be reliable, as multiple maintenance periods will be required in the lifetime of a power plant. For example, it was estimated that about $11$ to $13$ replacements of the first wall would be required in the lifetime of the ARIES-CS power plant \citep{el-guebaly_2008}. In addition, the MS has to be fast, as a longer maintenance period makes the power plant less financially attractive --- a study by \citet{ward_2005}, where the PROCESS code \citep{kovari_2014,kovari_2016} was applied to study a tokamak fusion power plant, showed that the cost of electricity would scale as $\text{coe}\sim A^{-0.6}$, where $A$ is defined as the fraction of time where the power plant is operational. While this study focused on a tokamak power plant, it still shows an important correlation between the fusion power plant availability, and its cost of electricity. Usually, power plant availability is assumed to be of the order of $85\%$ \citep{el-guebaly_2008}. In addition to impacting the cost of electricity produced by a fusion power plant, its accessibility might also have consequences on the tritium economy --- as identified by \citet{abdou_2021}, tritium production is paused during maintenance, but the amount of tritium in the blanket decreases by radioactive decay. 

Different MSs for stellarators have been discussed in the literature. Recently, \citet{warmer_2024} compared qualitatively four different schemes. The first one considers only the use of vertical access ports, as for the DEMO tokamak \citep{bachmann_2022}. The second relies on both vertical and horizontal access ports, while the third considers large vertical access ports, accessible by translating one or more large coils around the device. Finally, the last considered remote MS assumes splitting the stellarator in different sections that can be moved radially. The qualitative comparison by \citet{warmer_2024} lead to no clear winner, and further studies on each scheme viability were deemed necessary. It was however recognized that, in general, remote maintenance in a stellarator is expected to be more challenging than in a tokamak, as stellarators typically have more coils, their shaping can reduce port sizes, and the plasma chamber shaping complicates the replacement of internal components. 

The freedom in three-dimensional shaping of both the plasma and the coils can however also be an advantage, as one could promote accessibility in a stellarator optimization, as it is shown in this paper in the context of a port-based MS. Other approaches, such as segment-based MSs, are not discussed. Note that accessibility also depends on coil topologies, as discussed by \citet{ku_2010}. In this paper, we only consider modular and windowpane coils, and do not consider helical coils \citep{yamaguchi_2021}. It is worth mentioning here that large access ports are not only important for reactor accessibility, but also for experimental stellarators. Often, heating systems such as neutral beam heating (NBI), or diagnostics, require large access ports to the plasma. For example, the W7-X coils limit the size and shape of the NBI heating system duct, ultimately limiting the NBI pulse length \citep{rust_2011}.


Arguably, the choice of MS will heavily depend on the design of the stellarator --- conversely, the stellarator design will depend on the choice of the MS. It is therefore important to develop tools to optimize for good plasma accessibility for each possible MS, and study the trade-off between the MS, other plasma objectives and engineering constraints, such that the right MS can be chosen for the right application. Earlier work on stellarator accessibility focused on the engineering challenges of designing MS for a specific reactor design. For the HELIAS-5B design, a MS involving an in-vessel crane that would bring plasma-facing components to various ports was considered \citep{schauer_2013a}. In the case of the ARIES-CS stellarator, a port-based MS was considered initially \citep{brown_2015}. In a second study, coils were optimized under the constraint to be straight on the outboard side of the device using the COILOPT++ code \citep{gates_2017}, allowing segments of the vessel to be moved radially to gain access to the device. These studies provided the best possible MS given a stellarator design; only recently were considered the reverse problem, were stellarators were optimized provided a MS. For example, \citet{lion_2023} used the PROCESS code to explore the trade-off between vertical access ports and other engineering constraints and physics objectives. To the authors knowledge, there are however no prior attempts to explicitly target access port size in a stellarator optimization.

Among the large parameter space of possible stellarators \citep{giuliani_2022a,giuliani_2023,giuliani_2023a}, some devices have good accessibility, indicating that it may be possible to design accessible stellarator reactors. Whether or not a device is accessible can be hard to assess --- here we loosely define accessible as devices with large space between coils. Good examples are the Columbia Non-Neutral Torus (CNT) \citep{pedersen_2004b}, or more recently the Compact Stellarator with Simple Coils (CSSC) \citep{yu_2022}, or the Columbia Stellarator eXperiment (CSX) \citep{baillod_2025a} currently being designed at Columbia University. None of these examples include all of the metrics required for a reactor though. Another example is the Eos neutron source stellarator design \citep{gates_2025b,swanson_2025,kruger_2025}, where an array of non-encircling windowpane coils shapes the plasma, while toroidal field coils are planar and leave large openings between them. 


Large access ports are typically obtained indirectly by satisfying other engineering constraints --- for example by using only a small number of coils, or by enforcing toroidal field coils to be planar as in the Eos design. In this paper, we propose an alternative approach where large access ports are explicitly targeted in the objective function. We therefore focus on finding stellarators that allow large access ports between the coils without choosing a small number of coils, nor constraining their geometry. A new representation for access ports is developed (section \ref{sec.math_description}), and several penalty functions to maximize the plasma accessibility assuming a port-based accessibility scheme are derived (section \ref{sec:port_size_eval}). This novel approach is explored and leveraged to design three dimensional modular coils that allow for large access ports (section \ref{sec:stage_II}). We explore both vertical and horizontal access ports, and their trade-offs with the magnetic field quality. \blue{A current potential argument is used to explain the resulting port locations}. Finally, section \ref{sec:wps} explores the possibility to use windowpane coils to improve configurations and obtain coil sets that allow for large access ports and good magnetic field quality.

\section{Mathematical description of a port}\label{sec.math_description}
\begin{figure}
    \centering
    \includegraphics{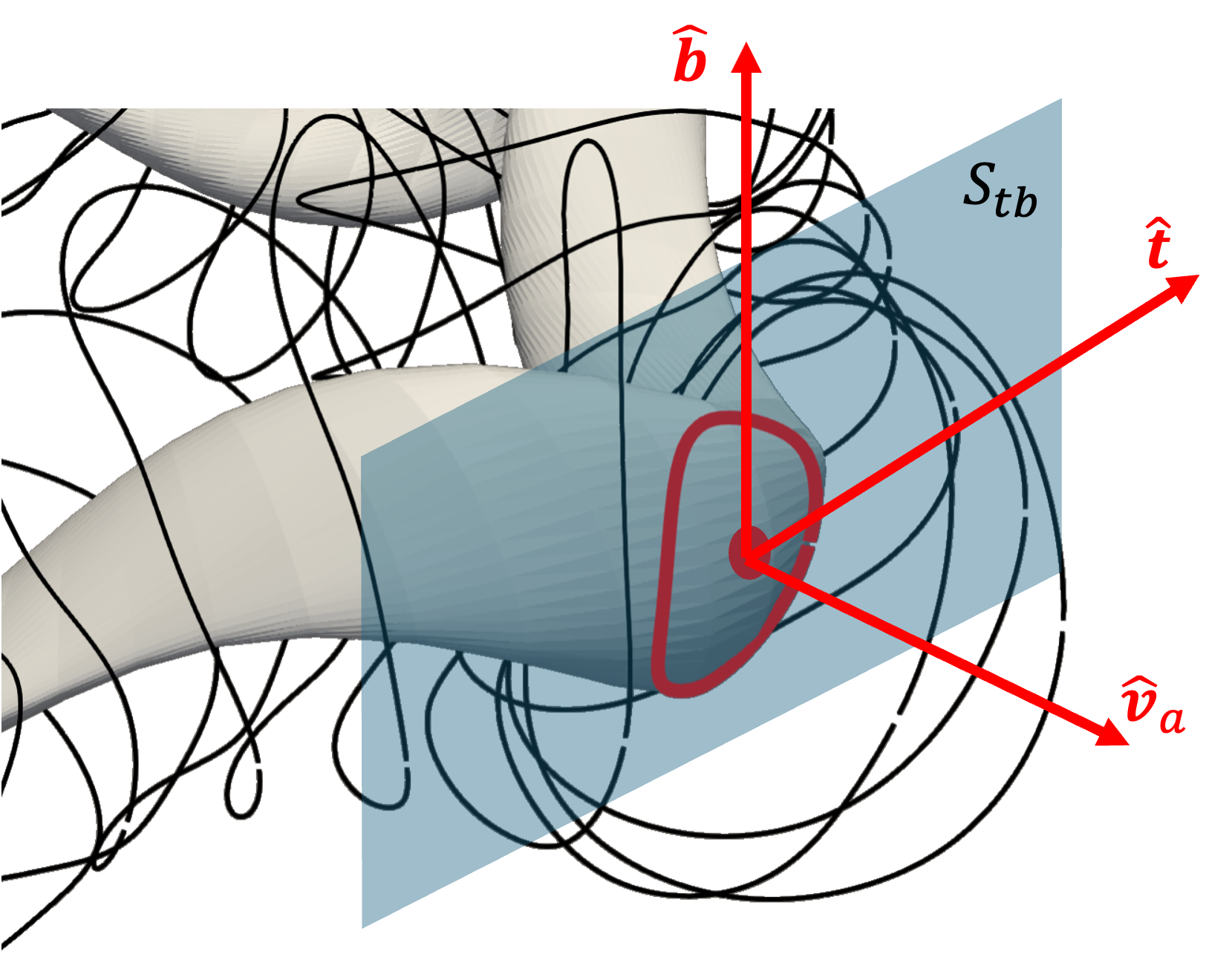}
    \caption{Port coordinate system $(\hat{\mathbf{v}}_a, \hat{\mathbf{t}}, \hat{\mathbf{b}})$. White: plasma boundary $\Gamma$. The black curves are coils, while the red curve on $\Gamma$ is the port boundary $C_p$. The coordinate system $(\hat{\mathbf{v}}_a, \hat{\mathbf{t}}, \hat{\mathbf{b}})$ is centered at $\mathbf{x}_0$, and the basis vectors $(\hat{\mathbf{t}}, \hat{\mathbf{b}})$ span the tangent plane $S_{tb}$, in blue.}
    \label{fig.port_coordinate}
\end{figure}

We define an access port as a region on the plasma boundary that can be reached along a specified access direction (see Figure \ref{fig.port_coordinate}). A point on the plasma boundary is considered accessible if there exists a line of sight in the given access direction that does not intersect any immovable components of the reactor, such as the external coils that confine the plasma. Each point within the access port has an associated line of sight, and the collection of these lines of sight forms a three-dimensional volume known as the port duct, which is depicted in green in the top panels of Figure \ref{fig:largest_port_solution}. The boundary of the access port, referred to as the port edge, is the closed curve formed by the intersection of the port duct with the plasma boundary. Note that the access port should ideally be constrained to be on the surface defined by the plasma facing components, \textit{i.e.} the first wall. The access port is however chosen here to be constrained on the plasma boundary, as in general the design of the first wall is unknown in stellarator optimization. The optimization method described in this paper can however be straight-forwardly adapted if the shape of the first wall is known. In the following section, we provide a more rigorous mathematical formulation of the port description.


Working in cylindrical coordinates $(R,\phi,Z)$, we describe the plasma boundary $\Gamma(\theta,\phi): [0,2\pi]\times[0,2\pi] \rightarrow\mathbb{R}^3$ as a double Fourier series,
\begin{align}
    R(\theta,\phi) &= \sum_{m=0}^M\sum_{n=-N}^N R_{mn}\cos(m\theta-nN_f\phi)
    \label{eq.doublefourier1}
    \\
    Z(\theta,\phi) &= \sum_{m=0}^M\sum_{n=-N}^N Z_{mn}\sin(m\theta-nN_f\phi),
    \label{eq.doublefourier2}
\end{align}
where $\theta$ is the poloidal angle, $(M, N)$ is the largest poloidal and toroidal Fourier mode used to describe the plasma boundary respectively, and $N_f$ is the number of field period. For convenience, we pack all the $\{R_{mn}, Z_{mn}\}$ harmonics into a single array $\mathbf{d}_s$. Here, stellarator symmetry is assumed \citep{dewar_1998}.

We describe a port $C_p$ as a closed curve $\mathbf{x}_p(l):[0,1]\rightarrow \mathbb{R}^3$ on the plasma boundary $\Gamma$, where $l$ parameterized the curve $\mathbf{x}_p$, to which we associate a unit vector $\hat{\mathbf{v}}_a$, that represent the access direction to the port. To constrain mathematically the curve $\mathbf{x}_p$ to be on $\Gamma$, we write
\begin{align}
    \theta_p(l) &= \theta_{0} + \sum_{i=1}^{Q} \theta_{cn}\cos(2\pi nl) + \theta_{sn}\sin(2\pi nl), \label{eq.curvecws1}\\
    \phi_p(l)   &= \phi_{0}   + \sum_{i=1}^{Q} \phi_{cn}\cos(2\pi nl)   + \phi_{sn}\sin(2\pi nl),\label{eq.curvecws2}
\end{align}
where $Q$ is the largest Fourier mode of both $\theta_p(l)$ and $\phi_p(l)$, and we define the curve $\mathbf{x}_p(l)$ as 
\begin{equation}
    \mathbf{x}_p(l) = \Gamma(\theta_p(l),\phi_p(l)).
\end{equation}
In this paper, it was found that only a few Fourier modes were necessary to describe a port. Larger values of $Q$ are possible, but care has to be taken to enforce the port to remain convex, and not intersect itself. To avoid such issues, we keep $Q$ small, $Q=1,2$. All the Fourier modes of $\mathbf{x}_p$ are packed in a single vector,
\begin{equation}
    \mathbf{d}_p=\{ \theta_{cn},\theta_{sn}, \phi_{cn}, \phi_{sn}\}_{n=1,\ldots,Q}.
\end{equation}
The vector $\mathbf{d}_p$ contains then all the degrees of freedom of the curve $\mathbf{x}_p(l)$. For completeness, we note here that the geometry of the curve $\mathbf{x}_p$ depends also on the surface geometry, \textit{i.e.} $\mathbf{x}_p=\mathbf{x}_p(l;\mathbf{d}_p,\mathbf{d}_s)$. 

Consider now the centroid of $C_p$, thereafter named $\mathbf{x}_0$. We define the plane $S_{tb}(\mathbf{x}_0)$ as the plane perpendicular to the access direction $\mathbf{v}_a$, and which contains $\mathbf{x}_0$. One can construct two perpendicular unit vectors, $(\hat{\mathbf{t}}, \hat{\mathbf{b}})$, that spans the plane $S_{tb}$. This defines an orthogonal coordinate system $(\hat{\mathbf{v}}_a, \hat{\mathbf{t}}, \hat{\mathbf{b}})$, centered at $\mathbf{x_0}$ (Figure \ref{fig.port_coordinate}). As an example, consider a vertical access port. In this particular case, we would chose $\hat{\mathbf{v}}_a$ as $\hat{\mathbf{z}}$, and $(\hat{\mathbf{t}},\hat{\mathbf{b}})=(\hat{\mathbf{x}},\hat{\mathbf{y}})$, with $(\hat{\mathbf{x}},\hat{\mathbf{y}},\hat{\mathbf{z}})$ the Cartesian coordinate system. A radial access port, on the other hand, would imply that $\hat{\mathbf{v}}_a$ is the cylindrical radial basis vector $\hat{\mathbf{r}}$, and $(\hat{\mathbf{t}},\hat{\mathbf{b}})=(\hat{\phi},\hat{\mathbf{z}})$.

In what follows, we will describe a point $\mathbf{y}$ as \textit{behind} $\mathbf{x}_p(l)$ if $(\mathbf{y}-\mathbf{x}_p(l))\cdot\hat{\mathbf{v}}_a<0$, and \textit{in front of} $\mathbf{x}_p(l)$ if $(\mathbf{y}-\mathbf{x}_p(l))\cdot\hat{\mathbf{v}}_a>0$. To ensure accessibility to the port, no immovable coils should be in front of a point $\mathbf{x}_p(l)$ if $\|\mathbf{x}_p(l)-\mathbf{y}\|_{tb}<\epsilon_{cp}$, where $\|\cdot\|_{tb}$ denotes the Euclidean distance in the $S_{tb}$ plane, and $
\epsilon_{cp}$ is a threshold distance. Coils that can be moved without being sectioned in multiple segments, for example windowpane coils, are not considered when evaluating the accessibility of the port, as they can be removed during maintenance. 

We will now consider two different optimizations, applied to two different configurations; the so-called "precise QA" and "precise QH" found by \citet{landreman_2022a}. In the first optimization, we seek to find the largest port size, given some fixed plasma boundary and a fixed set of coils. In the second, we seek to reproduce a given plasma boundary by optimizing a set of coils while ensuring the largest possible port size. The trade-off between the port size and the field quality will be explored by looking at their Pareto front. Comparisons will be made with coils found without the inclusion of a port. For the precise QA configuration, the coils found by \citet{wechsung_2022b}, while the coils found by \citet{wiedman_2023} will be used for comparison in the precise QH configuration.

All optimizations consider a vacuum field (no pressure nor currents in the plasma). This assumption is not critical to the optimizations presented hereafter; the same algorithm could in principle be applied to the design of coils for finite $\beta$ equilibria. The port representation, as well as the objective functions introduced below, are implemented in simsopt \citep{landreman_2021}, a python framework for stellarator optimization. We use the Broyden–Fletcher–Goldfarb–Shanno (BFGS) algorithm \citep{liu_1989} from the \emph{scipy.optimize} Python package \citep{virtanen_2020} to drive the optimization. 

\section{Port size evaluation} \label{sec:port_size_eval}
\subsection{Target function and penalties}
To increase the machine accessibility, we search for a curve $\mathbf{x}_p$ that maximizes the port size, defined as the area enclosed by $\mathbf{x}_p$ projected on $S_{tb}$, while ensuring that there are no coils in front of $\mathbf{x}_p$. Given a point $\mathbf{x}(l) = x_n\hat{\mathbf{v}}_a + x_t\hat{\mathbf{t}} + x_b\hat{\mathbf{b}}$, we evaluate the port size with,
\begin{equation}
    A_{\text{port}}(\mathbf{d}_p,\mathbf{d}_s) = \frac{1}{2}\oint_{C_p} \left(x_t\frac{\partial x_b}{\partial l}+x_b\frac{\partial x_t}{\partial l}\right) dl. \label{eq.port_area}
\end{equation}
Note that in the results presented thereafter, the port area is normalized by the product of the plasma surface minor radius and major radius, \textit{i.e.} $A_{\text{port}}/aR_0$. This is a natural normalization as the plasma surface area scales as $S_\Gamma\sim 4\pi^2 R_0 a$. As a comparison point, the ARIES-CS design, with a major radius of $7.75$ m and minor radius of $1.7$ m, was designed with a port of $1.85\times 3.85=7.12\ \text{m}^2$, which gives a normalized port area of $A_{\text{port}}/aR_0 = 0.54$. While the normalized port area of ARIES-CS might be small in comparison to some of the results presented thereafter, it is important to keep in mind that the size of access ports in ARIES-CS were not explicitly maximized in its optimization; the port sizes reported here were found \textit{a posteriori}. Furthermore, the ARIES-CS design considered many additional engineering and physics objectives and constraints, which unavoidably competed with the port size. These additional engineering constraints are however outside the scope of the present paper. 

To find the largest port, we seek to minimize the target function
\begin{equation}
    \begin{split}
    f_{\text{port}}(\mathbf{d}_p,\mathbf{d}_c,\mathbf{d}_s) = -w_{\text{area}}A_{\text{port}}(\mathbf{d}_p,\mathbf{d}_s) + w_{cp}J_{cpdist}(\mathbf{d}_p,\mathbf{d}_s,\mathbf{d}_c) \\
    + w_{ffp}  J_{ffp}(\mathbf{d}_p,\mathbf{d}_s) + w_{\text{arc}} J_{\text{arc}}(\mathbf{d}_p), \label{eq.fport}
    \end{split}
\end{equation}
where $\{ w_{\text{area}}, w_{cp},  w_{ffp},  w_{\text{arc}}\}$ are user-supplied scalar weights, $J_{cpdist}(\mathbf{d}_p,\mathbf{d}_s,\mathbf{d}_c)$ penalizes ports that have coils blocking its access, $\mathbf{d}_c$ are the degrees of freedom determining the coils position, $J_{ffp}(\mathbf{d}_p,\mathbf{d}_s)$ penalizes ports that wrap around the plasma boundary, and $J_{\text{arc}}(\mathbf{d}_p)$ penalizes ports that are parametrized with varying arc-length. We detail each penalty function below.

We consider a set of $N_{\text{coils}}$, represented by the curves $\{C_i\}_{i=\{1,\ldots,N_{\text{coils}}\}}$, with $\mathbf{x}^i_c(l)\in C_i$, $l\in[0,1)$. The coil-to-port distance penalty function is defined as
\begin{equation}
    J_{cpdist} = \sum_{i=1}^{N_{\text{coils}}} \oint_{C_i}\oint_{C_p}  \max\left[(\mathbf{x}_{c}^i(l_i)-\mathbf{x}_{p}(l_p))\cdot\mathbf{\hat{v}}_a, 0\right]^2 \max[d_{th}-|\mathbf{x}_{p}(l_p)-\mathbf{x}_{c}^i(l_i)|_{tb}, 0]^2dl_idl_p, \label{eq.cpdist}
\end{equation}
with $d_{th}$ a threshold distance, and $|\cdot|_{tb}$ the Euclidean distance in the $S_{tb}$ plane. In Eq.(\ref{eq.cpdist}), the first maximum function is non-zero only if there is a point $\mathbf{x}_{c}^i(l_i)$ in front of $\mathbf{x}_{p}(l_p)$, while the second maximum is non-zero only if the distance in the $S_{tb}$ plane between both points is smaller than the threshold distance. Both maximum functions are squared, to ensure the differentiability of $J_{cpdist}$. To summarize, $J_{cpdist}$ is non-zero if and only if there is at least one point from a coil that is in front of \textit{and} close to the port edge.

The forward-facing port penalty forces the port not to wrap around the plasma boundary. It is defined as
\begin{equation}
    J_{ffp} = \oint_{C_p} \max[-\hat{\mathbf{n}}(l_p)\cdot\mathbf{\hat{v}}_a, 0]^2 dl_p, \label{eq.ffp}
\end{equation}
with $\hat{\mathbf{n}}(l_p)$ the unit vector normal to the plasma boundary at $\mathbf{x}_{p}(l_p)$. This function is non-zero if anywhere along $C_p$ the normal to the plasma boundary points in the opposite direction than the access direction.

Finally, we use the arc-length penalty implemented in simsopt to constraint the parameterization of $\mathbf{x}_p$. We represent $\mathbf{x}_p$ in real space by a set of points $\{\mathbf{x}_{p,i}\}_{i=\{1,\ldots,N_{p}\}}$. The arc-length penalty penalizes parameterizations of $\mathbf{x}_p$ that deviate too far from the constant arc-length parameterization. This is an important penalty; without it, the optimizer is free to change the parameterization to have large gaps between the discrete points $\mathbf{x}_{p,i}$, and positions the points in between coils, therefore ensuring that $J_{cpdist}=0$ even though $\mathbf{x}_p$ crosses the curve of the coil. It is defined as,
\begin{equation}
    J_{\text{arc}} = Var\left(\left\{\left|\frac{\partial \mathbf{x}_{p,i}}{\partial l}\right|\right\}_{i=1,\ldots,N_p}\right),
\end{equation}
\textit{i.e.} it penalizes the variance of $\mathbf{x}_p$ tangent vector. 

If some of the weights $\{w_{\text{area}},w_{ffp},w_{cp},w_{\text{arc}}\}$ are not carefully chosen, some penalties can nevertheless be violated, for example a port can get behind a coil. An additional post-processing diagnostic is therefore applied to evaluate if a coil gets in front of a port. This algorithm follows three steps: \textit{projecting}, \textit{segmenting}, and \textit{sorting}. In the \textit{projecting} step, the curves describing the coils are projected in the $(\hat{\mathbf{v}}_a,\hat{\mathbf{t}},\hat{\mathbf{b}})$ coordinate system. Then, the \textit{segmenting} stage splits the curves in multiple segments that are either in front or behind the port. Finally, the \textit{sorting} stage sorts whether some of the segments in front of the port are in the port duct or not. Each step is detailed below. 

We represent each coil $C_i$ by a finite number of points $\mathbf{x}^i_{c,k}= \mathbf{x}^i_c(l_k)$. In what follows, the coil index $i$ is dropped for readability. In the \textit{projecting} stage of the algorithm, we compute the components of $\mathbf{x}_{c,k}$ in the $(\hat{\mathbf{v}}_a,\hat{\mathbf{t}},\hat{\mathbf{b}})$ coordinate system (see Figure \ref{fig.port_coordinate}), \textit{i.e.} we write
\begin{equation}
    \mathbf{x}_{c,k} = v_k\hat{\mathbf{v}}_a + t_k\hat{\mathbf{t}} + b_k\hat{\mathbf{b}}.
\end{equation}

In the \textit{segmenting} stage, we sort the points $\mathbf{x}_{c,k}$ in two categories: those in front of the port, and those behind the port. This is done by comparing their $v_k$ coordinate with the nearest point on the port boundary, denoted by $\mathbf{x}_{p,k}^*$,
\begin{equation}
    \mathbf{x}_{p,k}^* = \argmin_j \|\mathbf{x}_{c,k}-\mathbf{x}^*_{p,j}\|^2 = v^*_k\hat{\mathbf{v}}_a + t^*_k\hat{\mathbf{t}} + b^*_k\hat{\mathbf{b}}.
\end{equation}
Points in front of the port then satisfy $v_k - v^*_k \geq 0$. Note that due to the curvature of the plasma boundary, some points $\mathbf{x}_{c,k}$ are behind the port centroid, but in front of part of the port boundary --- this is why the position of points $\mathbf{x}_{c,k}$ has to be compared to their nearest neighbor on the port boundary.

In the last step of the algorithm, we evaluate whether the points in front of the port are in its duct or not. To sort these points, their winding number with respect to the port curve are evaluated in the $S_{tb}$ plane; points with a winding number greater than zero are in the port duct. The winding number $\omega(\mathbf{x}_{c,k}, C_p)$ is evaluated with
\begin{equation}
    \omega(\mathbf{x}_{c,k}, C_p) = \oint_{C_p} \frac{\left[t_k-t^*_k\right]^2}{|\mathbf{x}_{c,k}-\mathbf{x}_p(l_p)|_{tb}} db - \frac{\left[b_k-b^*_k\right]^2}{|\mathbf{x}_{c,k}-\mathbf{x}_p(l_p)|_{tb}} dt, 
\end{equation}
with $d\mathbf{x} = \hat{\mathbf{v}}_a dv + \hat{\mathbf{b}} db + \hat{\mathbf{t}} dt$. If there is a point $x_{c,k}$ for which $\omega(\mathbf{x}_{c,k}, C_p)>0$, then there is a point inside the port duct; therefore
\begin{equation}
    \text{Number of points inside the port duct} = \sum_k \omega(\mathbf{x}_{c,k}, C_p).
\end{equation}
This post-processing algorithm finds the number of points that belongs to a coil in the port duct. If this number is greater than zero, then a coil intersects the port duct --- the optimization failed, and the weights have to be adapted. This post-processing algorithm is more robust that the penalty described in Eq.(\ref{eq.cpdist}), as it picks up any point inside the port duct, even if it is far from the port boundary. It is however not suitable as a penalty term in the objective function \eqref{eq.fport}, as the argument of the minima used in the segmenting stage is not differentiable with respect to the degrees of freedom.

\subsection{Optimizing port size with fixed coils}
In this first optimization, we consider the precise QH plasma boundary by \citet{landreman_2022a} with the coils recently published by \citet{wiedman_2023}, and the precise QA plasma boundary by \citet{landreman_2022a} with the coils obtained by \citep{wechsung_2022b}. We keep the plasma boundary and the coils fixed and seek the largest radial access port in the case of the QA device, and the largest vertical access port in the case of the QH configuration. More specifically, we fix $\{\mathbf{d}_s,\mathbf{d}_c\}$, and explore the parameter space spanned by $\{\mathbf{d}_p\}$. We set the weights to $w_{cp}=10^3$, $w_{ffp}=1$ and $w_{\text{arc}}=10^{-1}$. As a local optimization algorithm is used, the solution will depend on the initial condition. Here we select the initial condition that leads to the largest port size for visualization purposes; this initial port location is easily found visually by looking at the coil set and the plasma boundary. 

\begin{figure}
    \centering
    \hfill
    \includegraphics[width=.475\linewidth]{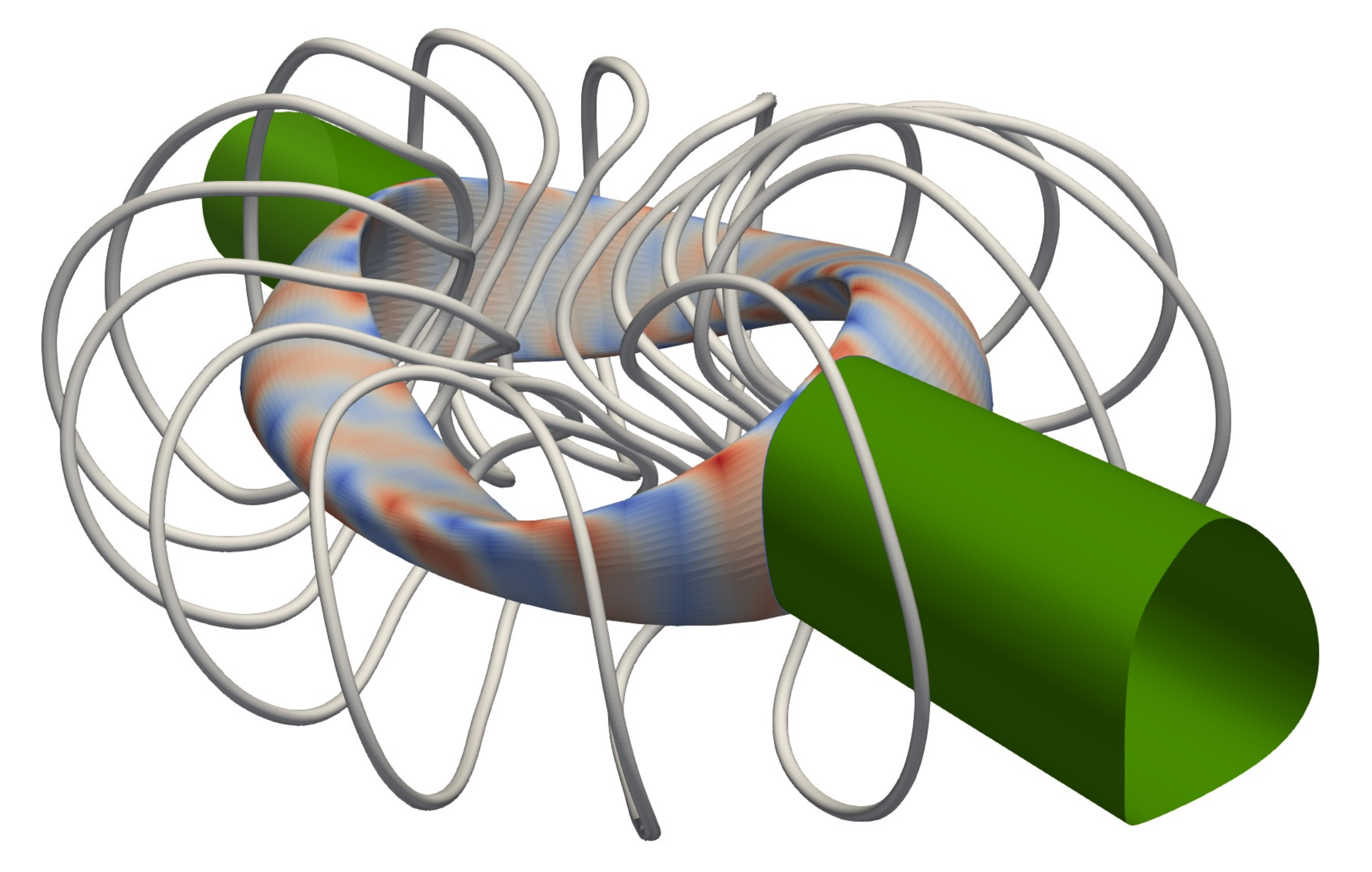}
    \hfill
    \includegraphics[width=.475\linewidth]{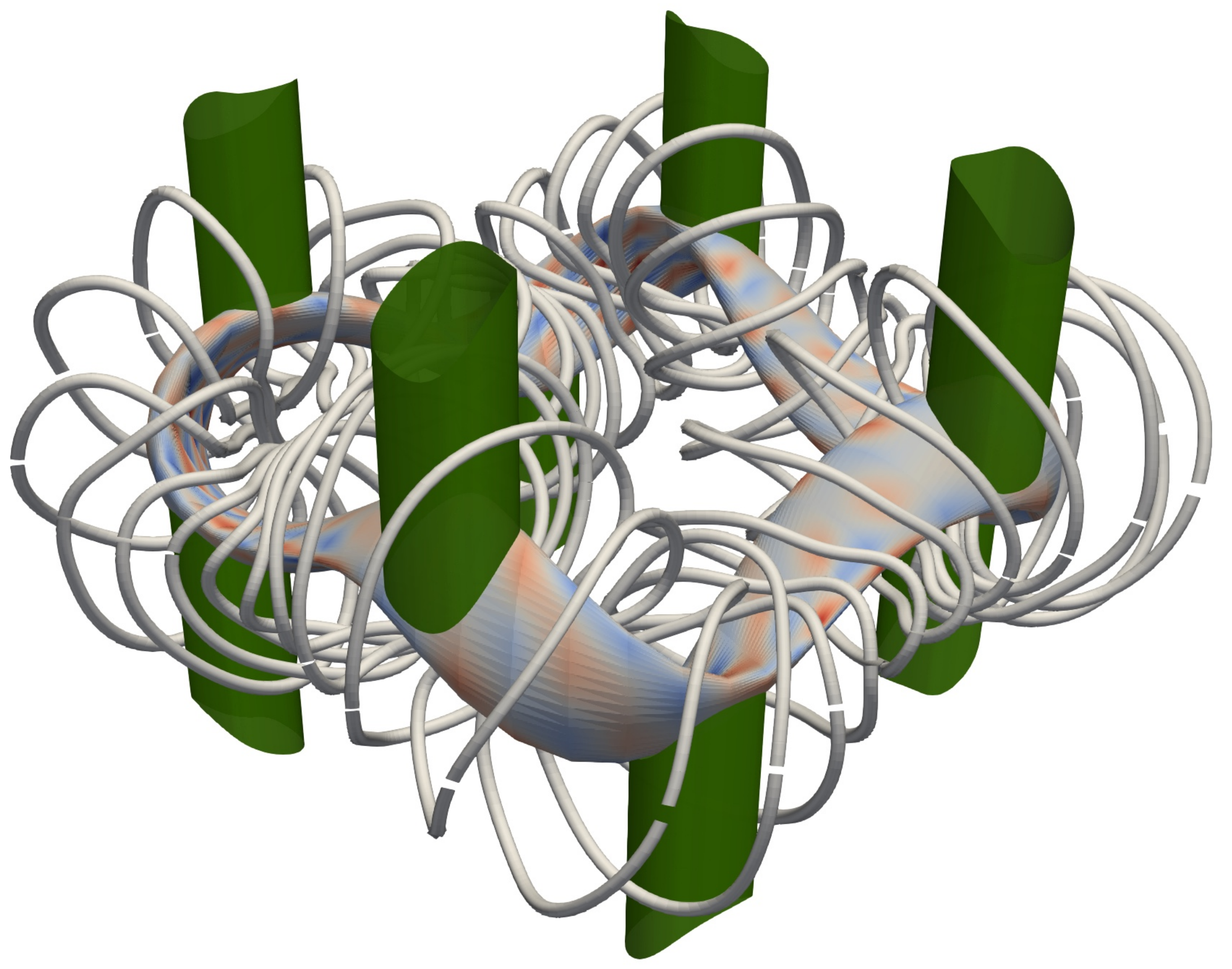}
    \hfill
    \\
    \hfill
    \includegraphics[width=.475\linewidth]{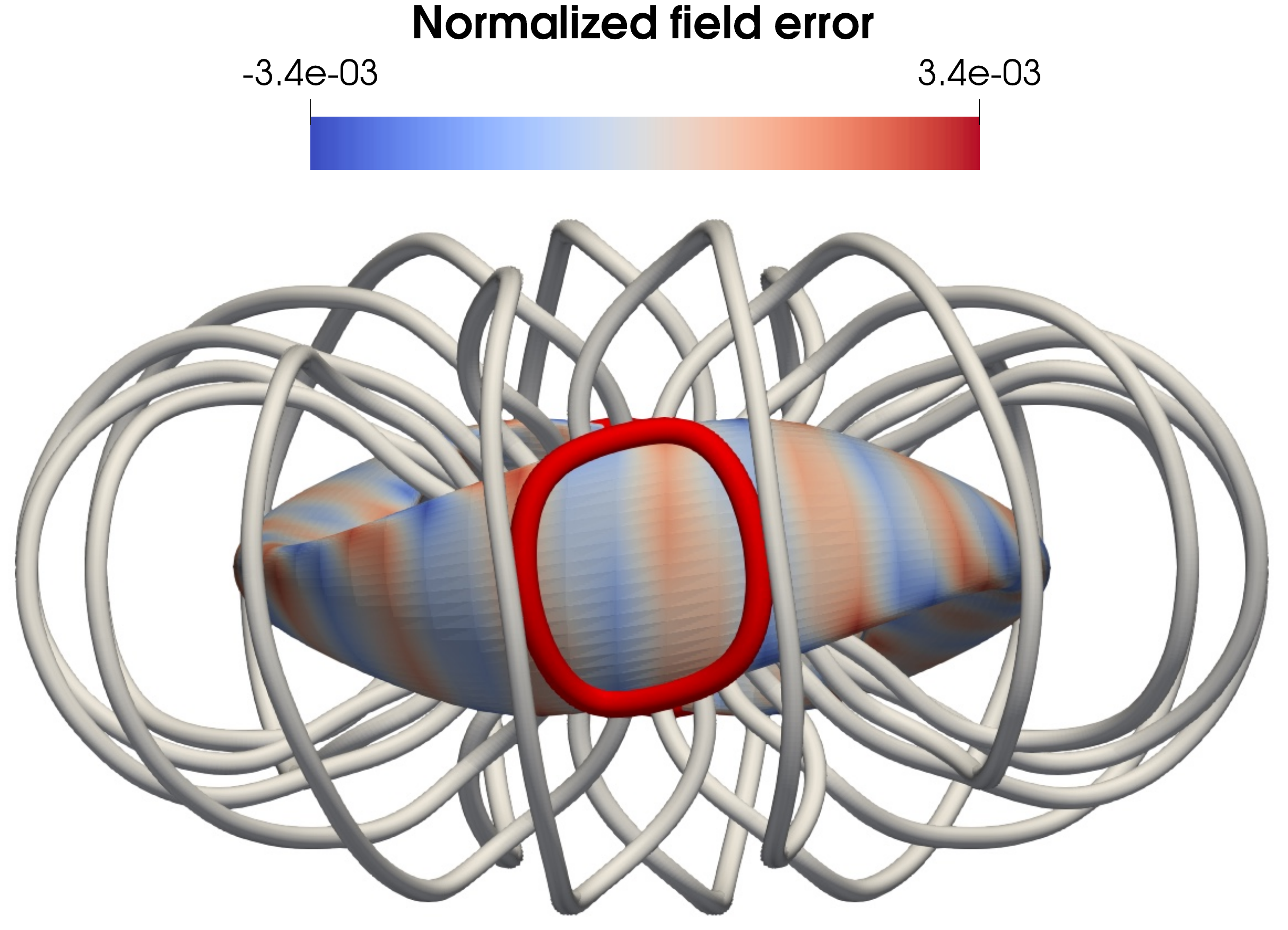}
    \hfill
    \includegraphics[width=.475\linewidth]{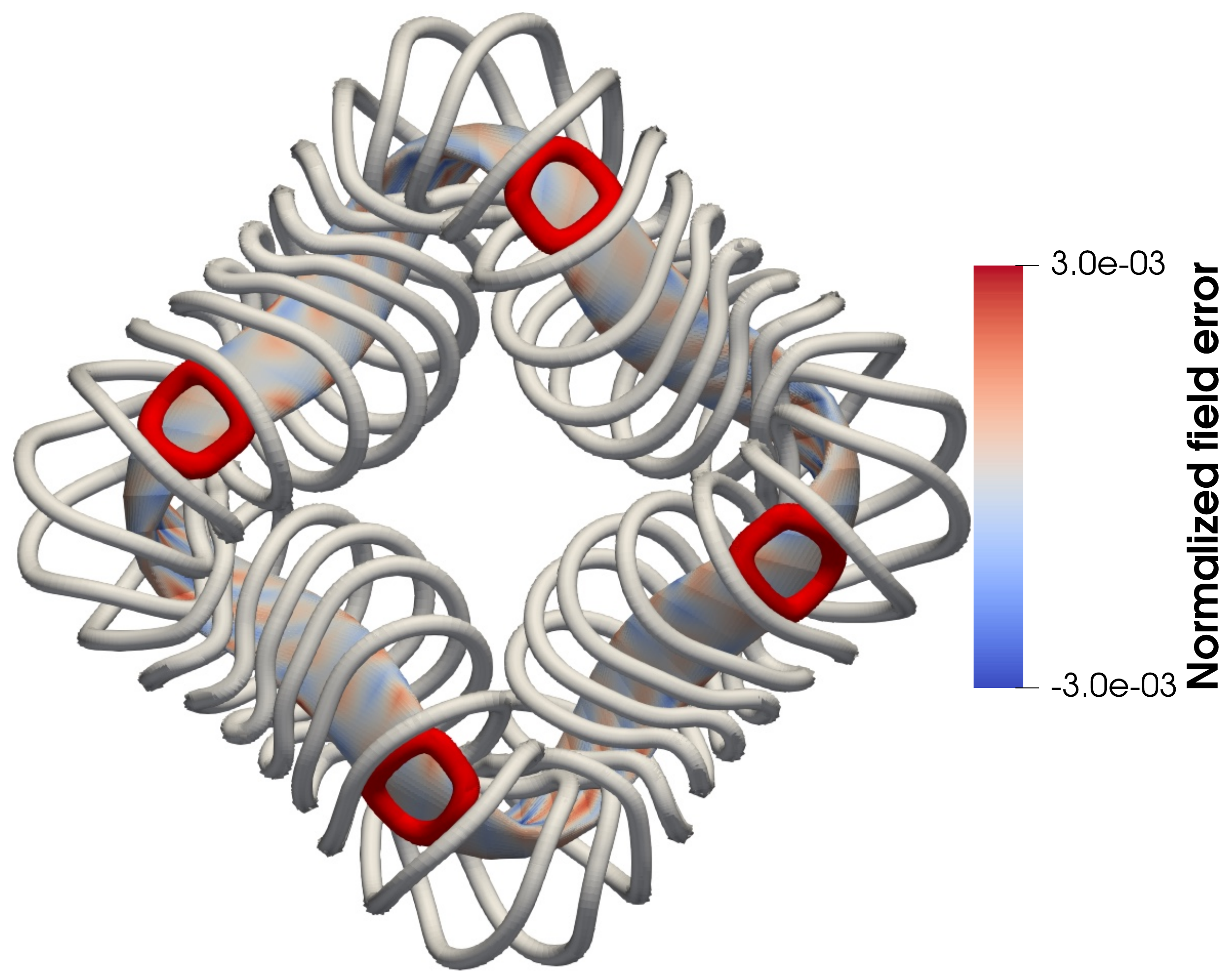}
    \hfill
    \caption{Configurations obtained by keeping the coils fixed; only the port shape was optimized to maximize its enclosed area. Precise QA configuration with radial access port (left) and precise QH configuration with vertical access port (right). Top: 3D visualization with port ducts in green and coils in gray. Bottom: point of view in the access direction. The coils (in gray) are the coils found by \citet{wechsung_2022b} (left) and \citet{wiedman_2023} (right). The colors on the plasma boundary indicate the normalized field error, \textit{i.e.} $\mathbf{B}\cdot\hat{\mathbf{n}}/B$. }
\label{fig:largest_port_solution}
\end{figure}

Figure \ref{fig:largest_port_solution} shows the three dimensional plot of the plasma boundary, the coils, and the access ports characterizing the extremum of Eq.(\ref{eq.fport}). As expected, it describes an access port as large as possible, that fits on the plasma boundary and in between coils. Here, the normalized port areas are $A_{\text{port}}/aR_0 = 2.16$ and $A_{\text{port}}/aR_0 = 0.61$ for the QA radial access port and the QH vertical access port respectively. Note that in the particular case of the radial access port for the precise QA configuration, the port is centered around the symmetry axis of the device; therefore, only one port per field period is obtained. Regarding the vertical access port obtained on the precise QH device, equivalent ports are obtained \textit{below} the device by symmetry, \textit{i.e.} with access direction $\hat{\mathbf{v}}_a=-\hat{\mathbf{z}}$. 

With these two simple examples, we showed that the minimization of $f_{\text{port}}$ leads to the largest port that fits on the plasma surface, and in between existing coils. In what follows, we focus on optimizations where both the coils and the ports are optimized at the same time; this is commonly known as a stage II optimization in the context of stellarator optimization, in contrast to stage I optimizations where a plasma boundary is sought.

\section{Coil optimization}\label{sec:stage_II}
We now optimize both the coils and the access port at the same time, and consider the plasma boundary as fixed. We look for coils that can generate a magnetic surface that matches as closely as possible the target plasma boundary, while allowing the largest possible access port. There is a natural competition between both objectives; the largest port is obtained when coils near the access port are deformed or translated away, which impacts the magnetic field generated by the coils. Here, the plasma boundary parameterized by $\mathbf{d}_s$ is fixed, while we explore the parameter space spanned by $\{\mathbf{I},\mathbf{d}_c,\mathbf{d}_p\}$ to maximize the port area and minimize the perpendicular component of the magnetic field to the target boundary, where $\mathbf{I}$ contains the degrees of freedom describing the currents in the coils. Note that in what follows we consider 5 and 4 coils per field period for the precise QH and precise QA configurations respectively, unless stated otherwise.

\subsection{Objective function}
We seek to minimize the objective
\begin{align}
    f(\mathbf{I},\mathbf{d}_c,\mathbf{d}_s,\mathbf{d}_p) &= f_{II}(\mathbf{I},\mathbf{d}_c,\mathbf{d}_s) + f_{\text{port}}(\mathbf{d}_c,\mathbf{d}_s,\mathbf{d}_p), \label{eq.stage_II_objective}
\end{align}
where the stage II objective function is defined by 
\begin{align}
    f_{II}(\mathbf{I},\mathbf{d}_c,\mathbf{d}_s) &= f_{\text{quadflux}}(\mathbf{I},\mathbf{d}_c,\mathbf{d}_s) &\text{Quadratic flux objective}\\
    &+  w_L J_L(\mathbf{d}_c) &\text{Coil length penalty}\\
    &+ w_{ccdist} J_{ccdist}(\mathbf{d}_c) &\text{Coil-coil distance penalty}\\
    &+ w_{csdist} J_{csdist}(\mathbf{d}_c,\mathbf{d}_s) &\text{Coil-surface distance penalty}\\
    &+ w_\kappa J_\kappa(\mathbf{d}_c) &\text{Maximum curvature penalty}\\
    &+ w_{\bar\kappa} J_{\bar\kappa}(\mathbf{d}_c) &\text{Mean curvature penalty}\\
    &+ w_{LN}J_{LN}(\mathbf{d}_c)&\text{Linking number penalty}\\
    &+ w_{\text{arc}}J_{\text{arc}}(\mathbf{d}_c). &\text{Arc-length penalty}
\end{align}
The scalars $\{w_L,w_{ccdist},w_{csdist},w_\kappa,w_{\bar\kappa},w_{LN}\}$ are user-supplied weights, and we use the quadratic flux target and the penalty functions as defined in \citet{wiedman_2023}. We write down their expressions for completeness below. The quadratic flux target function is,
\begin{equation}
    J_{\text{quadflux}} = \frac{1}{2} \iint_\Gamma \left(\frac{\mathbf{B}\cdot\hat{\mathbf{n}}}{|B|}\right)^2ds,
\end{equation}
where $\mathbf{B}$ is the magnetic field and $\hat{\mathbf{n}}$ the surface unit normal vector. We normalize by the magnetic field norm to avoid the trivial solution where $\mathbf{B}=0$. The first penalty penalizes the total length of the coils,
\begin{equation}
    J_{L} = \max\left[\left(\sum_{i=1}^{N_{\text{coils}}} L_i\right) - N_{\text{coils}}L_T, 0 \right]^2, \label{eq:length_penalty}
\end{equation}
where $L_i$ is the length of the $i^\text{th}$ coil, and $L_T$ is an averaged target length. By penalizing the total length instead of each coil length separately, we allow solutions where one coil is longer, which can be advantageous for the existence of large access ports. The coil-coil and coil-surface distance penalties penalize coils that get too close to each other and to the plasma surface respectively. These are defined as
\begin{align}
    J_{ccdist} &= \sum_{i=1}^{N_{\text{coils}}} \sum_{j > i}^{N_{\text{coils}}} \oint_{C_i}\oint_{C_j}\max\left[d_{c,th}-|\mathbf{x}^i_c-\mathbf{x}^j_c|, 0\right]^2dl_idl_j\\
    J_{csdist} &= \sum_{i=1}^{N_{\text{coils}}} \oint_{C_i}\iint_{\Gamma}\max\left[d_{s,th}-|\mathbf{x}^i_c-\mathbf{x}_s|, 0\right]^2dl_ids,
\end{align}
where $d_{c,th}$ and $d_{s,th}$ are threshold distances below which the coil-coil and coil-surface penalties apply respectively, and $\mathbf{x}^i_c\in C_i$ and $\mathbf{x}_s\in \Gamma$. These threshold distances are usually set by engineering considerations. The maximum and mean curvature penalties penalizes coils that have maximum and mean curvature respectively above a given threshold value. They are defined as
\begin{align}
    J_\kappa &= \frac{1}{2}\sum_{i=1}^{N_{\text{coils}}} \oint_{C_i} \max\left[\kappa(\mathbf{x}^i_c)-\kappa_{th}, 0\right]^2 dl_i\\
    J_{\bar\kappa} &= \sum_{i=1}^{N_{\text{coils}}} \max\left[ \frac{1}{L_i}\oint_{C_i} \kappa^2(\mathbf{x}^i_c)dl_i    -\bar{\kappa}^2_{th} ,0\right]^2
\end{align}
where $\kappa_{th}$ and $\bar{\kappa}_{th}$ are threshold values, and $\kappa(\mathbf{x}^i_c)$ is the curvature of the curve $C_i$ at $\mathbf{x}^i_c$. The last penalty penalizes curves that are interlocked, and is defined as 
\begin{equation}
    J_{LN}=\max\left(\sum_{i=1}^{N_{\text{coils}}}\sum_{j>i}^{N_{\text{coils}}}\left[\Theta(C_i,C_j)\right] - 0.1, 0\right)^2, \label{eq.jln}
\end{equation}
where $\Theta(C_i,C_j)$ evaluates the linking number between the i$^{th}$ and j$^{th}$ coil,
\begin{equation}
    \Theta(C_i,C_j) = \frac{1}{4\pi} \oint_{C_i}\oint_{C_j} \frac{\mathbf{x}^i_c(l_i)-\mathbf{x}^j_c(l_j)}{|\mathbf{x}^i_c(l_i)-\mathbf{x}^j_c(l_j)|^3} dl_idl_j. \label{eq.linking_number}
\end{equation}
The threshold value of $0.1$ is used in Eq.(\ref{eq.jln}) as the integral in Eq.(\ref{eq.linking_number}) is never exactly zero due to numerical approximations \citep{wiedman_2023}.

\begin{figure}
    \centering
    \begin{tikzpicture}
    \node (t1) at (-4.2,2) {\includegraphics[width=.5\linewidth]{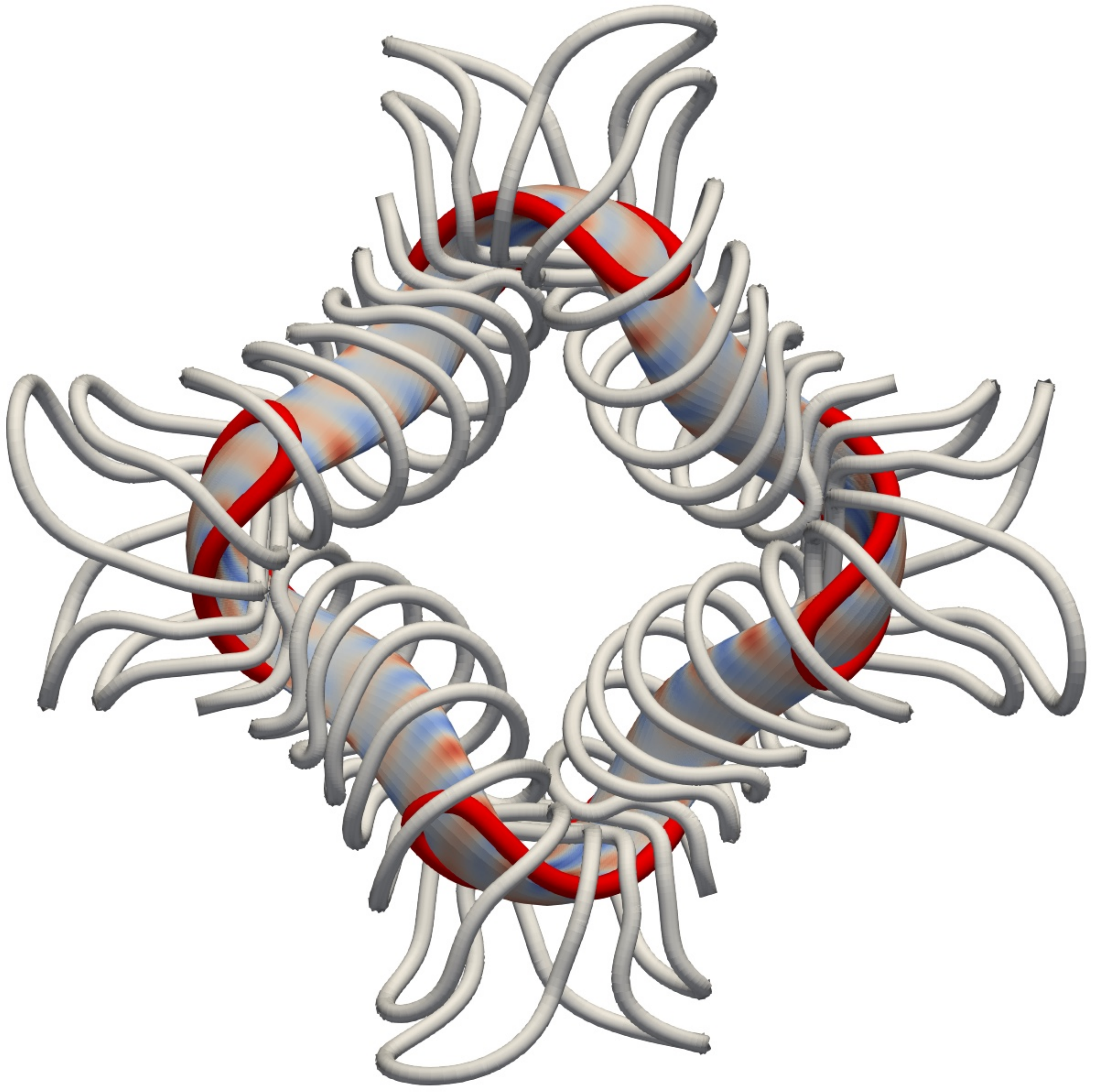}};
    \node(t3) at (4.2,2) {\includegraphics[width=.5\linewidth]{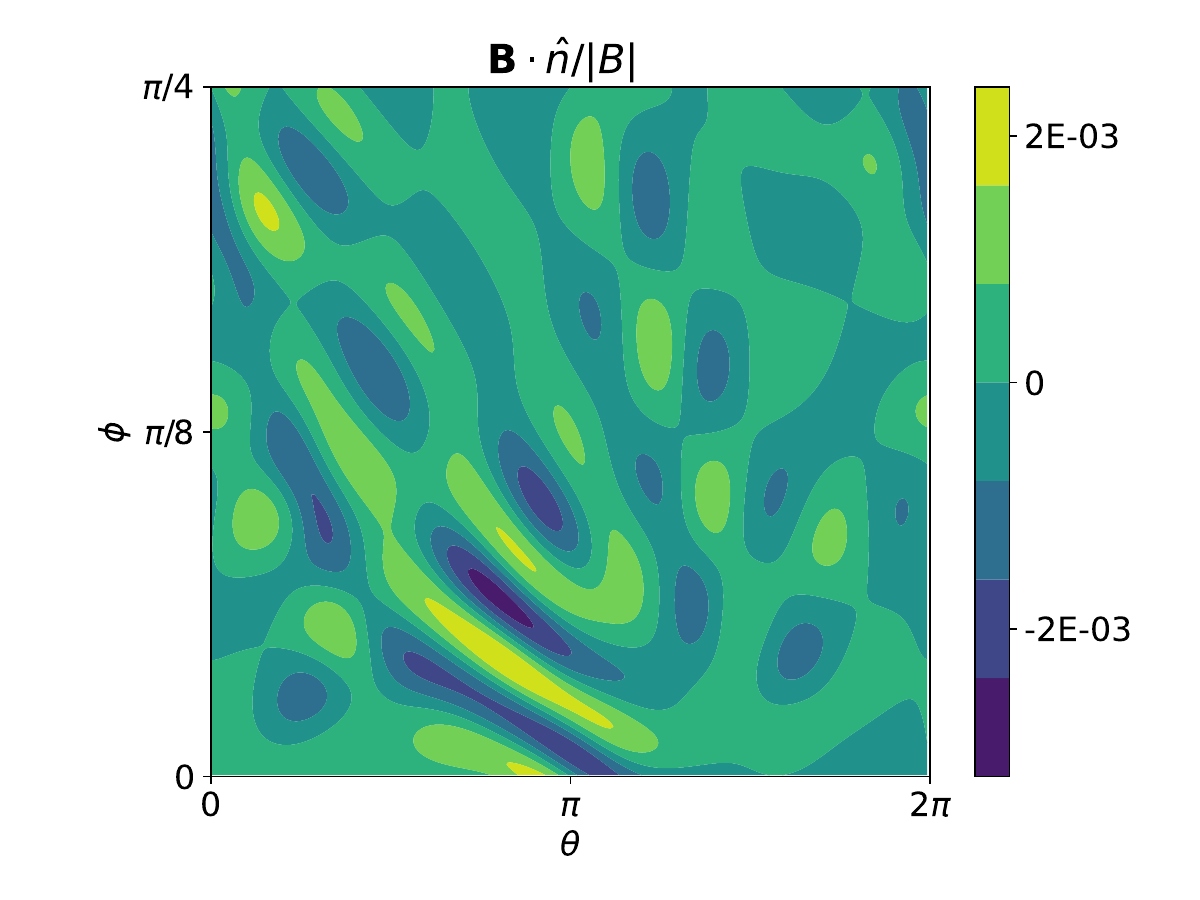}};
    \node (t2) at (0,-5) {\includegraphics[width=.7\linewidth]{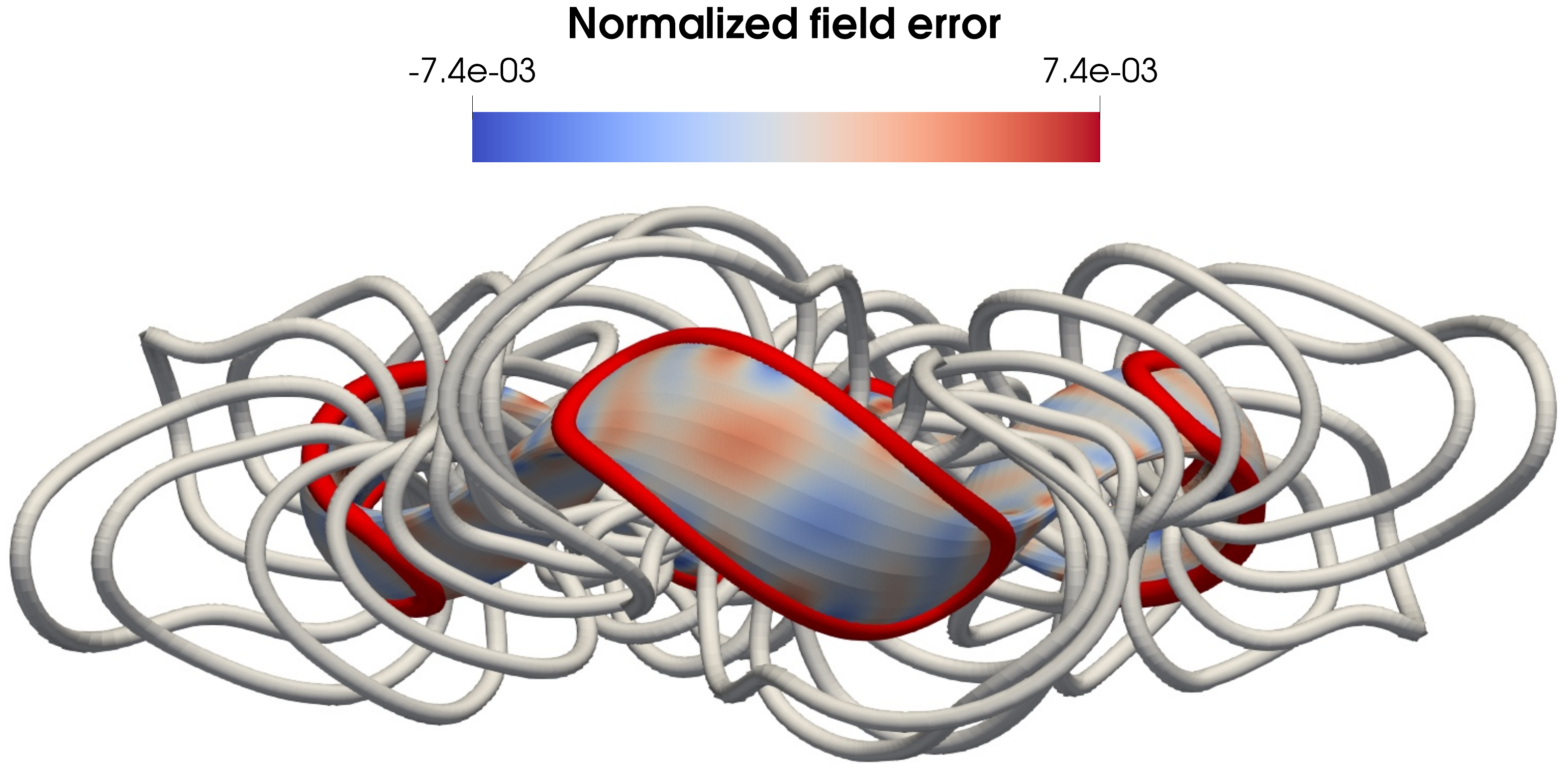}};
    \end{tikzpicture}
    \caption{Top view (top left) and side view (bottom) of the precise QH configuration with radial access port. Coils are plotted in gray, ports in red, and the colors on the plasma boundary are the normalized field error, \textit{i.e.} $\mathbf{B}\cdot\mathbf{n}/B$. Top right: Normalized field error over one half field period.}
    \label{fig:stage_II_radial_QH_3D_plot}
\end{figure}

\begin{figure}
    \centering
    \begin{tikzpicture}
        \node (t1) at (-4,0) {\includegraphics[height=5cm]{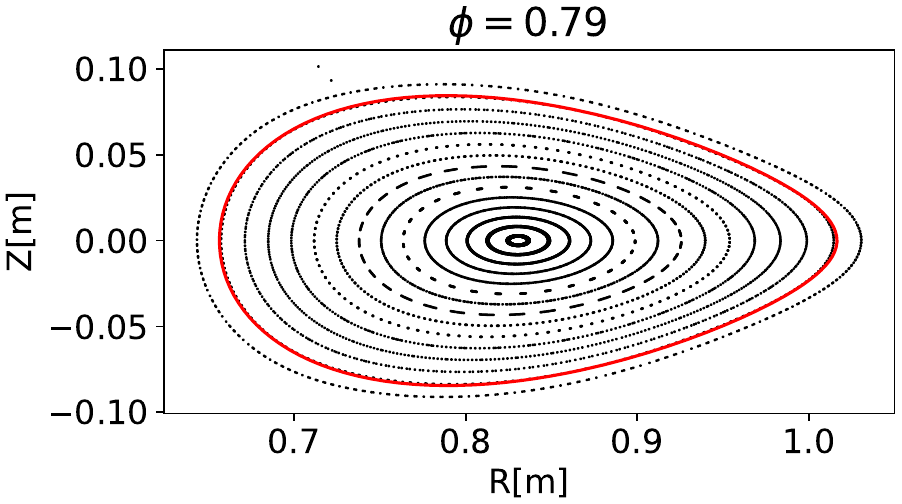}};
        \node (t1) at (4,0) {\includegraphics[height=7cm]{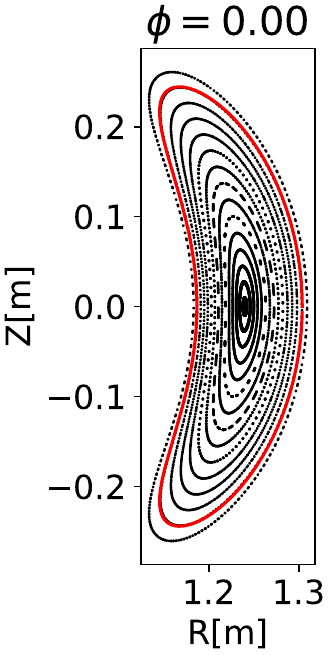}};
    \end{tikzpicture}
    \caption{Poincar\'e section (in black) at the triangular cross-section (left) and bean cross-section (right) for the precise QH configuration with radial access port. The red curve shows the target plasma boundary.}
    \label{fig:stage_II_radial_QH_field_quality}
\end{figure}

Results obtained for a radial access port in the case of the precise QH configuration are shown on Figure \ref{fig:stage_II_radial_QH_3D_plot}, where the coils, the port, and the plasma boundary are pictured. A port with normalized area of $A_{\text{port}}/aR_0 = 2.42$ is found. Interestingly, we see that the ports are located where the plasma boundary exhibits sharp bends. Coils around the port are deformed in order to clear the port duct. This results in elongated coils, that somewhat resemble famous potato chips. Figure \ref{fig:stage_II_radial_QH_field_quality} shows the normalized field error over one field period, and the Poincar\'e section at different toroidal sections. Perhaps surprisingly, we see that the plasma boundary is well recovered, despite the large deformation of the coils. This indicates that large radial access ports are possible to obtain, at least for this specific plasma boundary, without compromising the field quality. Similar radial access ports can be obtained in the case of the precise QA configuration (see Appendix \ref{app:raqa} for an example).

\begin{figure}
    \centering
    \begin{tikzpicture}
    \node (t1) at (-4,2) {\includegraphics[width=.5\linewidth]{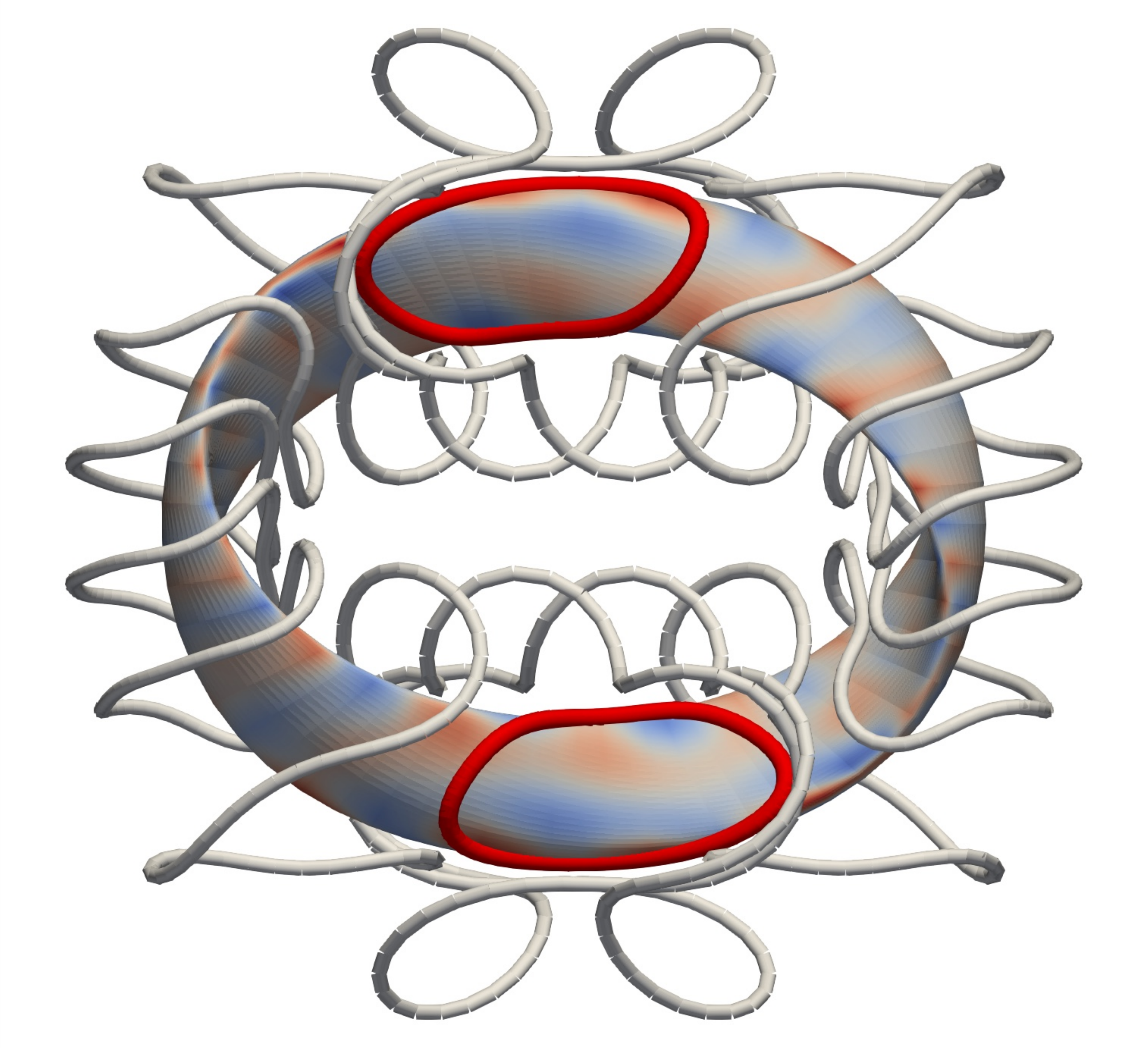}};
    \node (t3) at (4,2) {\includegraphics[width=.5\linewidth]{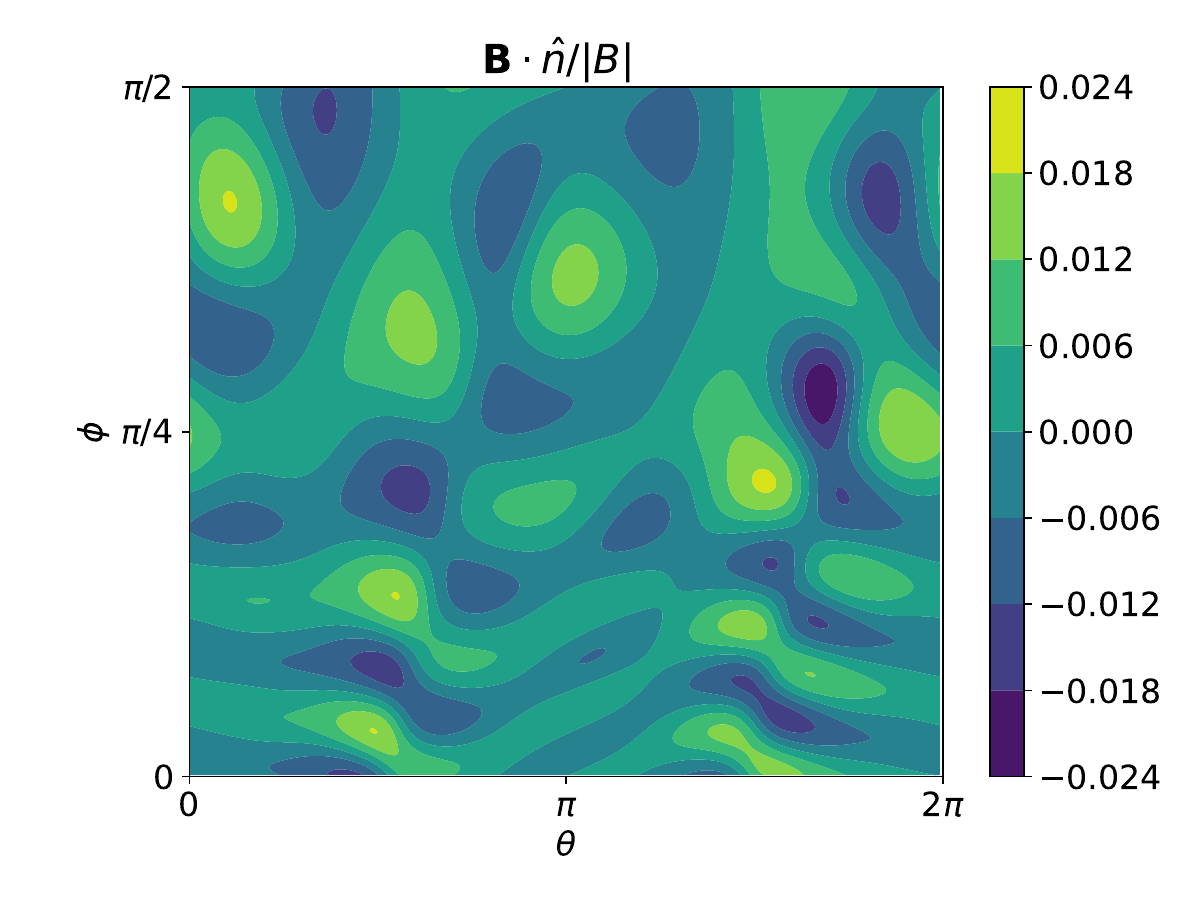}};
    \node (t2) at (0,-6.2) {\includegraphics[width=.8\linewidth]{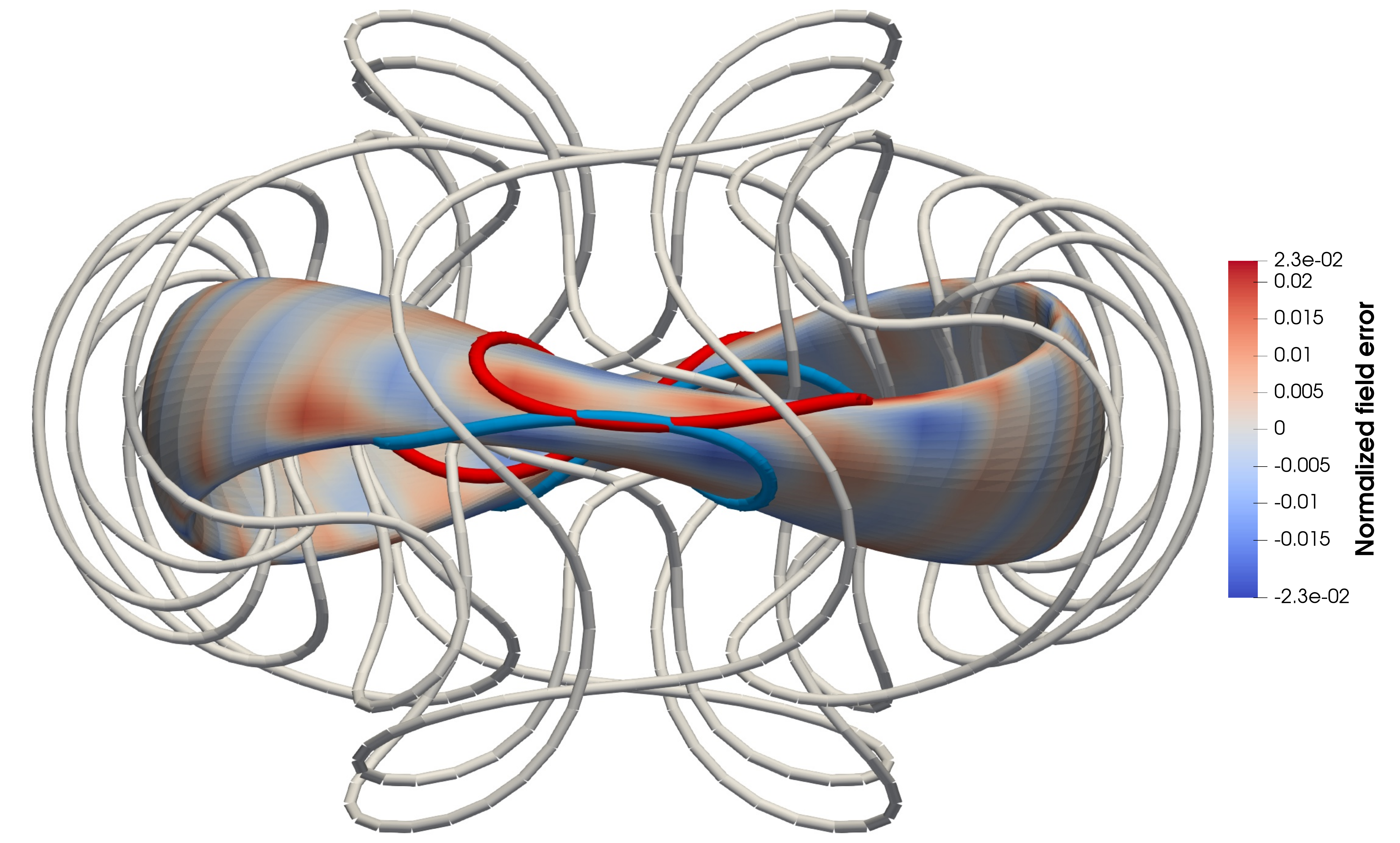}};
    \end{tikzpicture}
    \caption{Top view (top left) and side view (bottom) of the precise QA configuration with vertical access port. Coils are plotted in gray, upward facing ports in red, downward facing ports in blue, and the colors on the plasma boundary are the normalized field error, \textit{i.e.} $\mathbf{B}\cdot\mathbf{n}/B$. Top right: Normalized field error over one half field period.}
    \label{fig:stage_II_vertical_QA_3D_plot}
\end{figure}

\begin{figure}
    \centering
    \begin{tikzpicture}
    \node (t1) at (-4,0) {\includegraphics[width=0.625\linewidth]{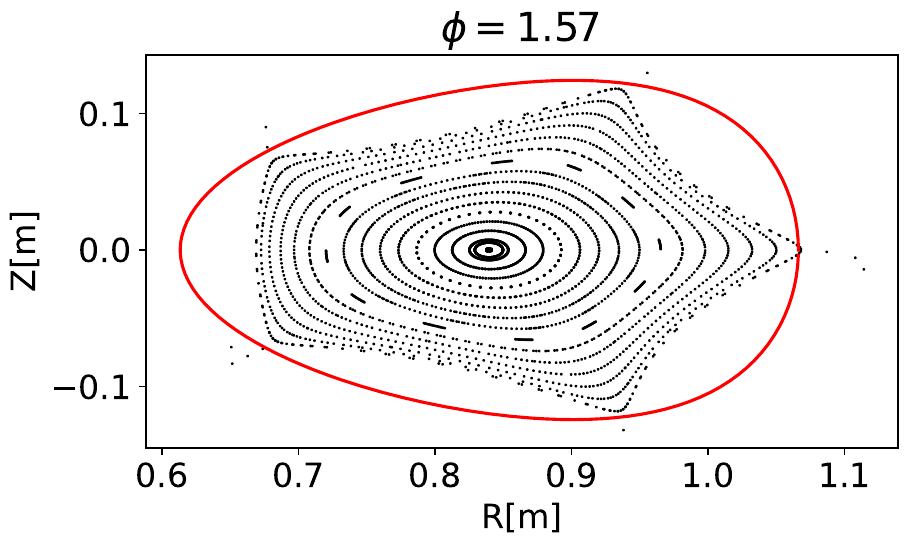}};    \node (t2) at (4,0) {\includegraphics[width=0.325\linewidth]{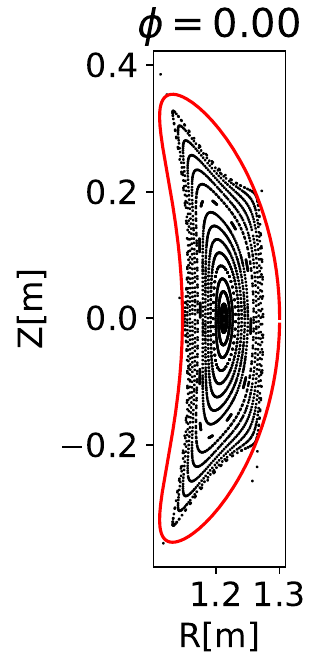}};
    \end{tikzpicture}
    \caption{Poincar\'e section (black) at the triangular cross-section (left) and bean cross-section (right) for the precise QA configuration with vertical access port. The red curve shows the target plasma boundary.}
    \label{fig:stage_II_vertical_QA_field_quality}
\end{figure}

Below is shown another example, where coils are deformed to obtain a vertical access port on the precise QA configuration.  Again, Figure \ref{fig:stage_II_vertical_QA_3D_plot} shows the 3D plot including the plasma boundary, coils and ports, and Figure \ref{fig:stage_II_vertical_QA_field_quality} shows the field error on the plasma boundary and the magnetic field Poincar\'e section at two different toroidal planes. This time, we observe than even though the port is of a similar size ($A_{\text{port}}/aR_0 = 2.37$) than the radial access port obtain for the QH configuration (see Figure \ref{fig:stage_II_radial_QH_3D_plot}), the coils deformation impacts the magnetic field more strongly, with $\max\mathbf{B}\cdot\hat{n}/B=2.4\%$. The plasma boundary is visually not recovered as a magnetic surface, and it appears that a large $m=5$ island chain formed at the plasma edge. The same is generally observed for the precise QH configuration (see Appendix \ref{app:vaqh} for an example) --- larger field errors are induced by a vertical access port than by a radial access port of the same size. Obviously, results with smaller ports and smaller normalized field error can be achieved by lowering the weight $w_{\text{area}}$; here this specific example was chosen to highlight the visually striking difference with the radial access port, despite having a port of similar size. 

\begin{figure}
    \centering
    \begin{tikzpicture}
    \node (t1) at (-4,2) {\includegraphics[width=.475\linewidth]{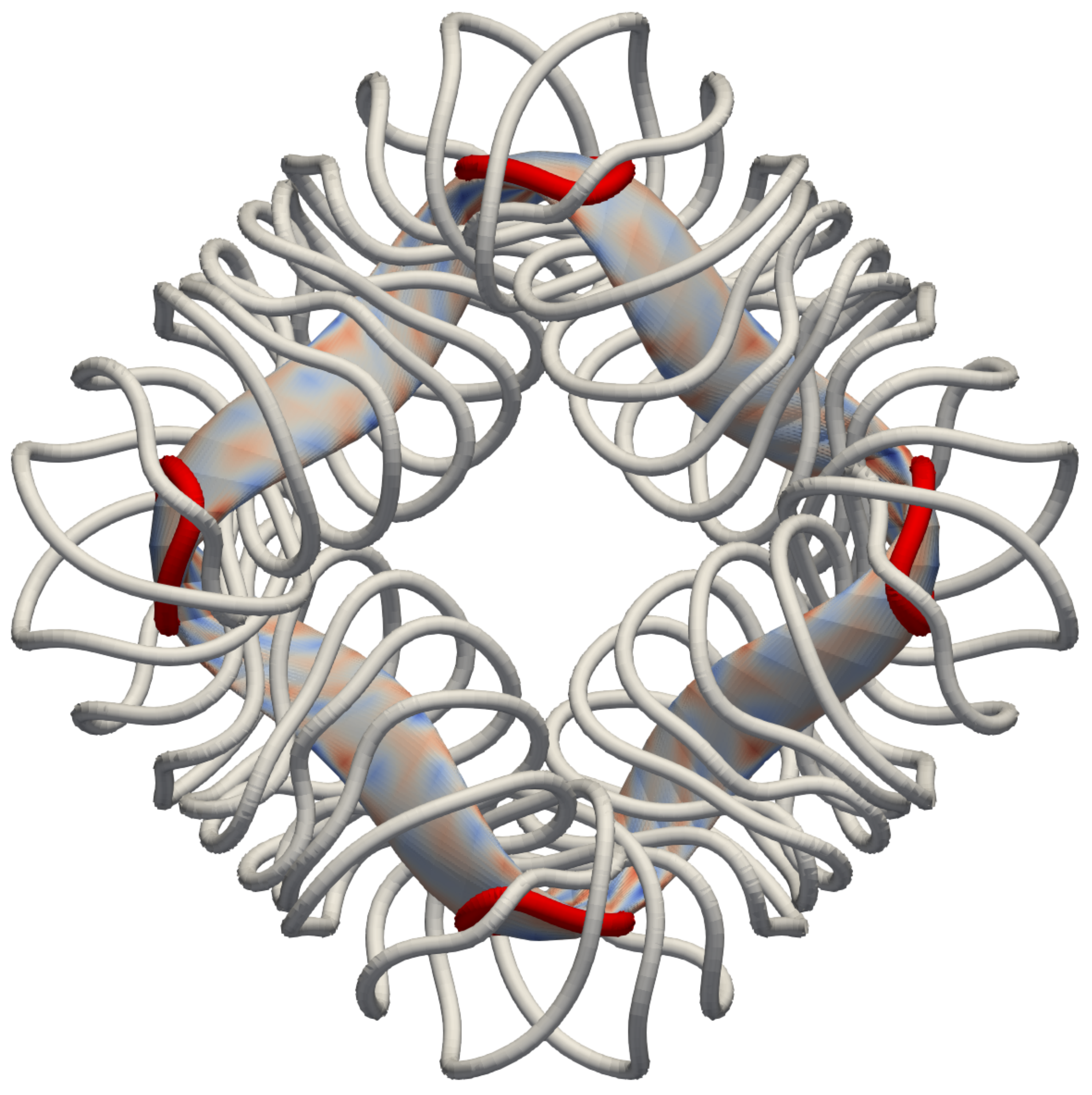}};
    \node (t3) at (4,2) {\includegraphics[width=.475\linewidth]{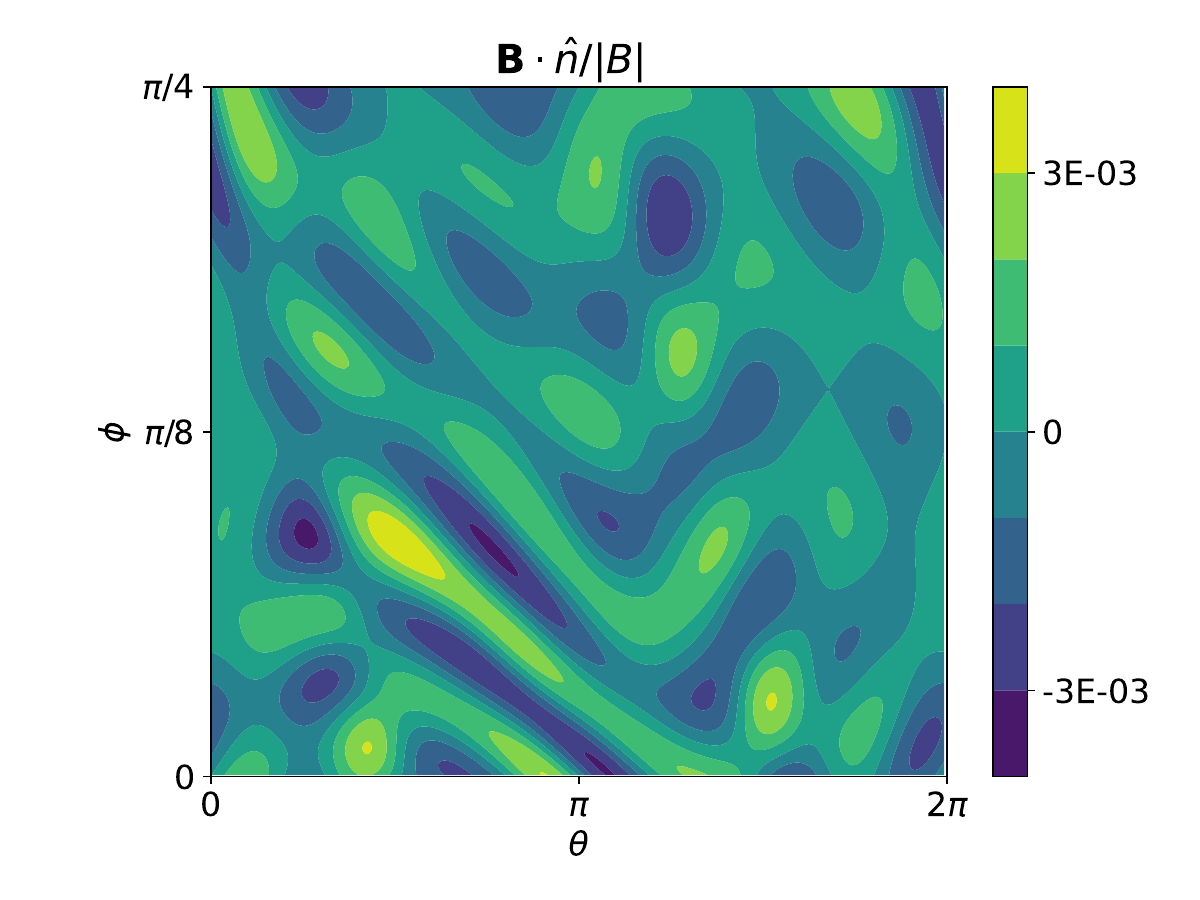}};
    \node (t2) at (0,-6) {\includegraphics[width=.8\linewidth]{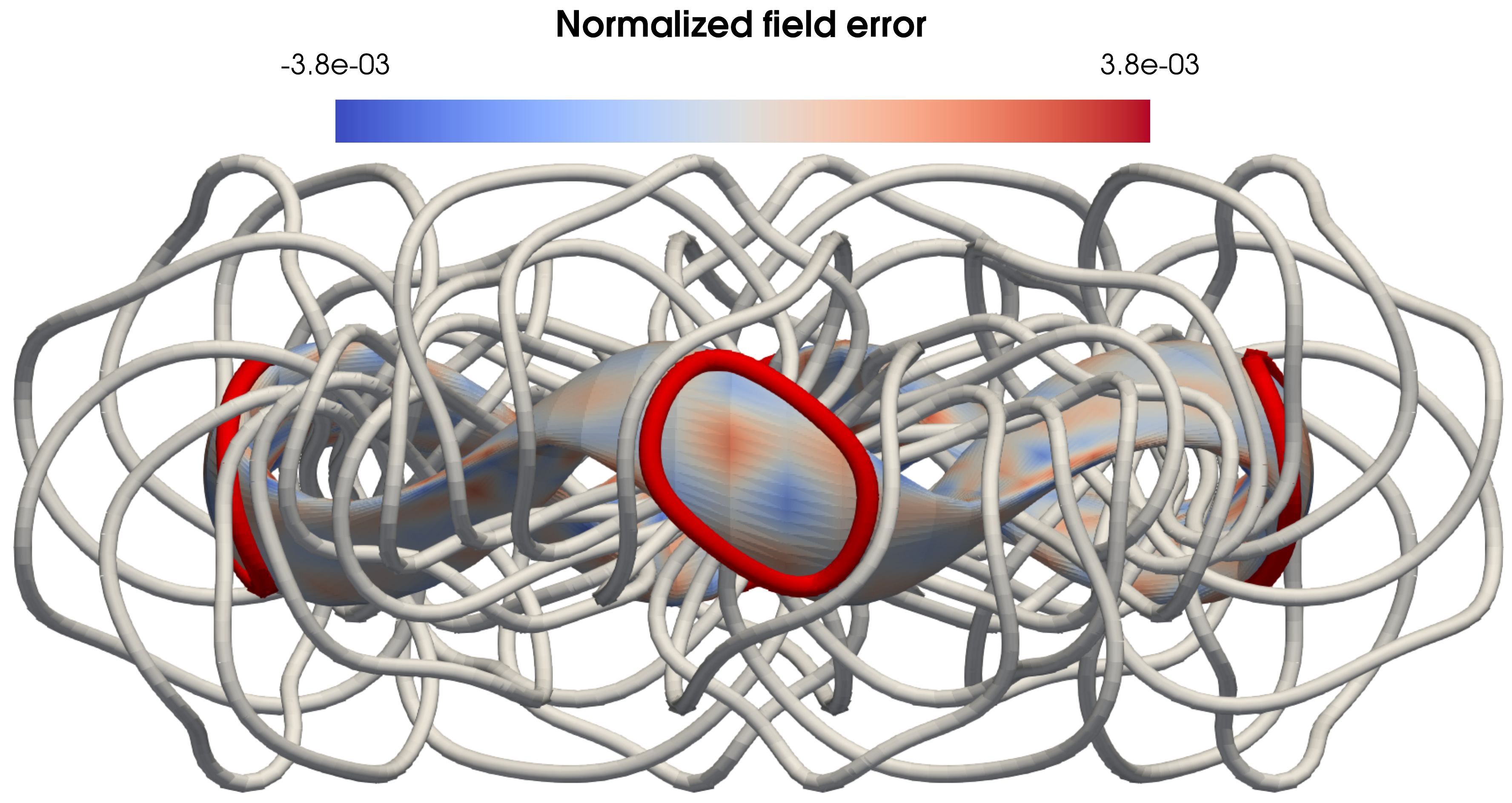}};
    \end{tikzpicture}
    \caption{Top view (top left) and side view (bottom) of the precise QH configuration with horizontal access port. In this example, coil lengths are individually constrained. Coils are plotted in gray, ports in red, and the colors on the plasma boundary are the normalized field error, \textit{i.e.} $\mathbf{B}\cdot\mathbf{n}/B$. Top right: Normalized field error over one half field period.}
    \label{fig:RAQH_LocalLengthConstraint_example}
\end{figure}

\begin{figure}
    \centering
    \begin{tikzpicture}
        \node (t1) at (-4,0) {\includegraphics[height=5cm]{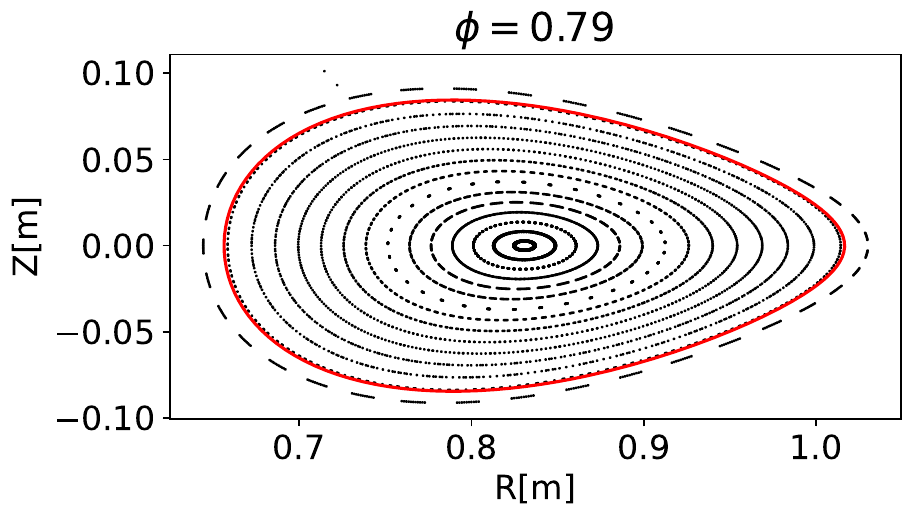}};
        \node (t2) at ( 4,0) {\includegraphics[height=7cm]{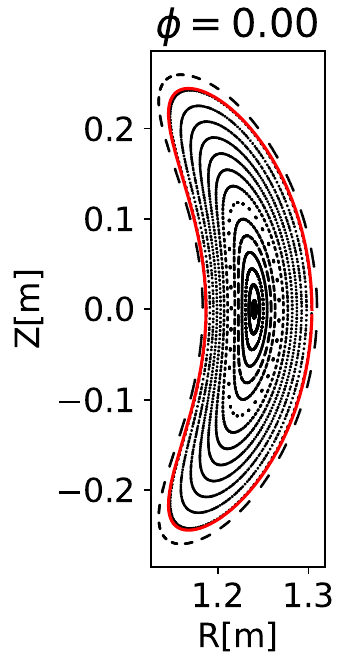}};
    \end{tikzpicture}
    \caption{Poincar\'e section (in black) at the triangular cross-section (left) and bean cross-section (right) for the precise QH configuration with radial access port, where coils length was constrained individually. The red curve shows the target plasma boundary.}
    \label{fig:stage_II_radial_QH_poincare_LocalLengthConstraint}
\end{figure}

Note that in the above examples, only the total length of the coils was penalized (see Eq.(\ref{eq:length_penalty})). This allowed some coils to get shorter, and others to get longer, such that large ports were possible --- see for example Figure \ref{fig:stage_II_radial_QH_3D_plot}.
The method described here to optimize for large access ports can, however, be used with other engineering constraints on the coils. 
As an example, we explore now the impact of constraining the length of each individual coil. This constraint may be of interest as having coils of similar dimensions may facilitate the design of a support structure. Effectively, we replace Eq.(\ref{eq:length_penalty}) by
\begin{equation}
    J_L = \sum_{i=1}^{N_{\text{coils}}}\max\left(L_i - L_T, 0\right)^2.
\end{equation}
This new set of constraints is obviously more constraining than the penalty on the total length of the coils. As a consequence, solutions with large ports are only achievable at the expense of field quality. In other words, there is a stronger trade-off between individual coil length and the access port size. An example is shown on Figure \ref{fig:RAQH_LocalLengthConstraint_example}. Again, we observe that a port is achievable by deforming coils next to it; however the normalized field error is about twice as large as in the case shown on Figure \ref{fig:stage_II_radial_QH_3D_plot}, for a normalized port of size of $A_{\text{port}}/aR_0=1.52$, about $37\%$ smaller! The coils nevertheless still achieve to reproduce the target boundary, as the Poincar\'e sections attest on Figure \ref{fig:stage_II_radial_QH_poincare_LocalLengthConstraint}.

\subsection{Field quality and port size trade-off}
Both optimizations shown above, \textit{i.e.} the precise QH with radial access port and the precise QA with vertical access port, are examples. As a local optimizer is used to minimize the objective function, the result depends on the initial guess, in particular the initial position of the port. In addition, the solution depends on the weights, especially the weight $w_{\text{area}}$ that determines the relative importance between the field quality and the port size in Eq.(\ref{eq.stage_II_objective}). To explore more solutions, we devise an algorithm that automatically selects weights related to penalty functions. Starting with small weights, a first optimization is executed. Then, each penalty is evaluated; for each unsatisfied penalty, its corresponding weight is increased, and the optimization is restarted from the initial guess. After a few iterations on the penalty weights, all penalties are satisfied, and the algorithm stops. With this algorithm, only the $w_{\text{area}}$ weight has to be chosen manually. 

To explore the trade-off between both objectives, we randomize the initial state, and the value of $w_{\text{port}}$, and run several hundreds of optimizations. Here modular coils are initialized as circular coils, where their radius, and position is picked randomly within some reasonable bounds. The access port is initialized as an ellipse in the $\theta-\phi$ plane. The initial guess is constructed such that all penalties are satisfied at the beginning of the optimization. 

For each obtained configuration, we then compare the normalized enclosed port area, as evaluated by Eq.(\ref{eq.port_area}), to the surface averaged error in the magnetic field, evaluated as
\begin{equation}
    \Delta B = \sqrt{\frac{1}{S_\Gamma}\iint_\Gamma\left(\frac{\mathbf{B}\cdot\hat{\mathbf{n}}}{B}\right)^2 dS},
\end{equation}
where $S_\Gamma=\iint_\Gamma dS$ is the plasma boundary surface. It should be noted that the surface averaged field error, $\Delta B$, while being useful as a simple scalar to grasp the overall field quality, does not necessarily correlate with the destruction of magnetic surfaces, or more broadly with a worsening of the confinement. Indeed, what really matters is to which mode of the normal magnetic field the plasma is sensitive to, and what is the amplitude of said mode \citep{zhu_ishw_2024}. Nevertheless, the averaged field error is a useful metric to measure how good the coils are to reproduce the target plasma shape, \textit{i.e.} how well does the coil set solve the stage II optimization problem.

\begin{figure}
    \centering
    \includegraphics[width=0.85\linewidth]{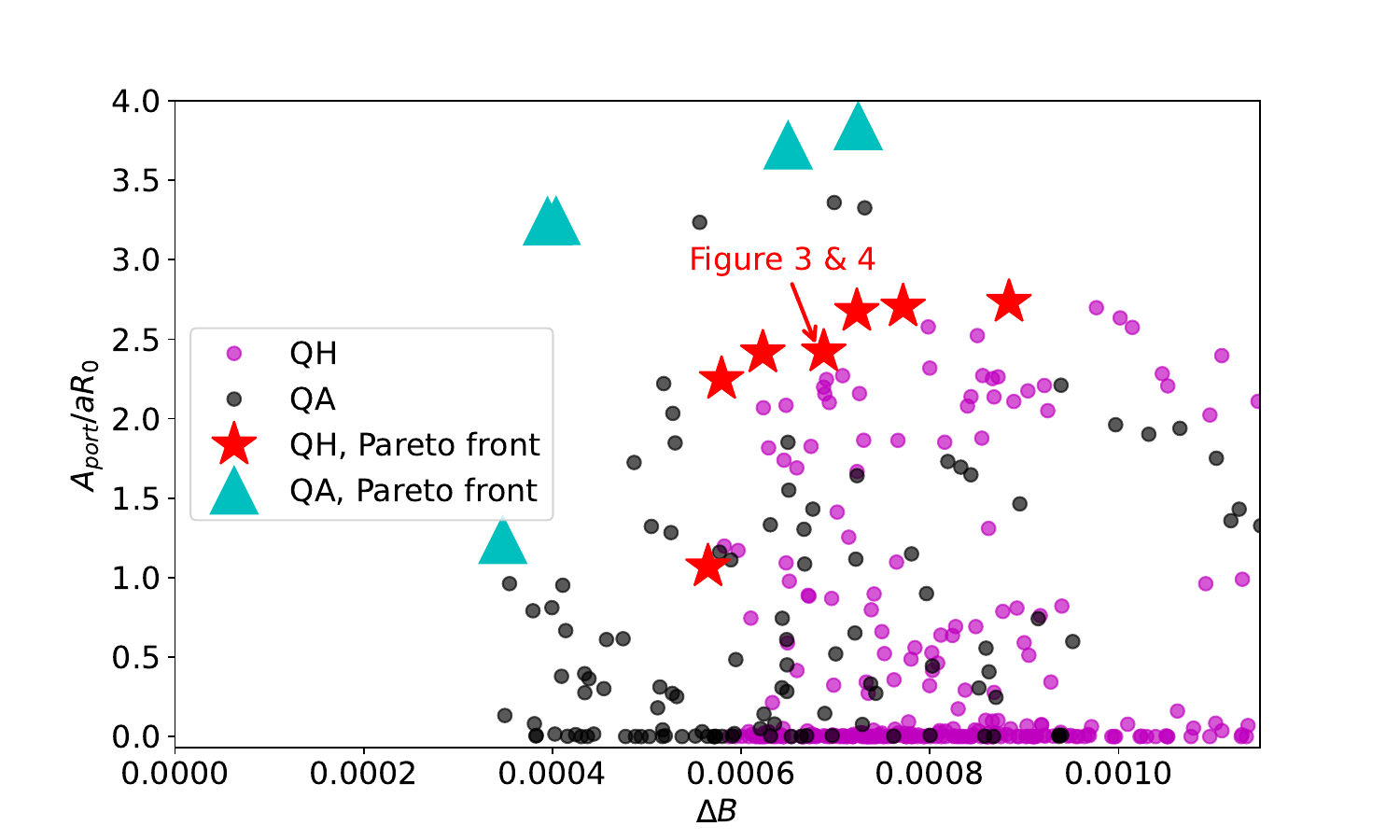} \\
    \includegraphics[width=0.85\linewidth]{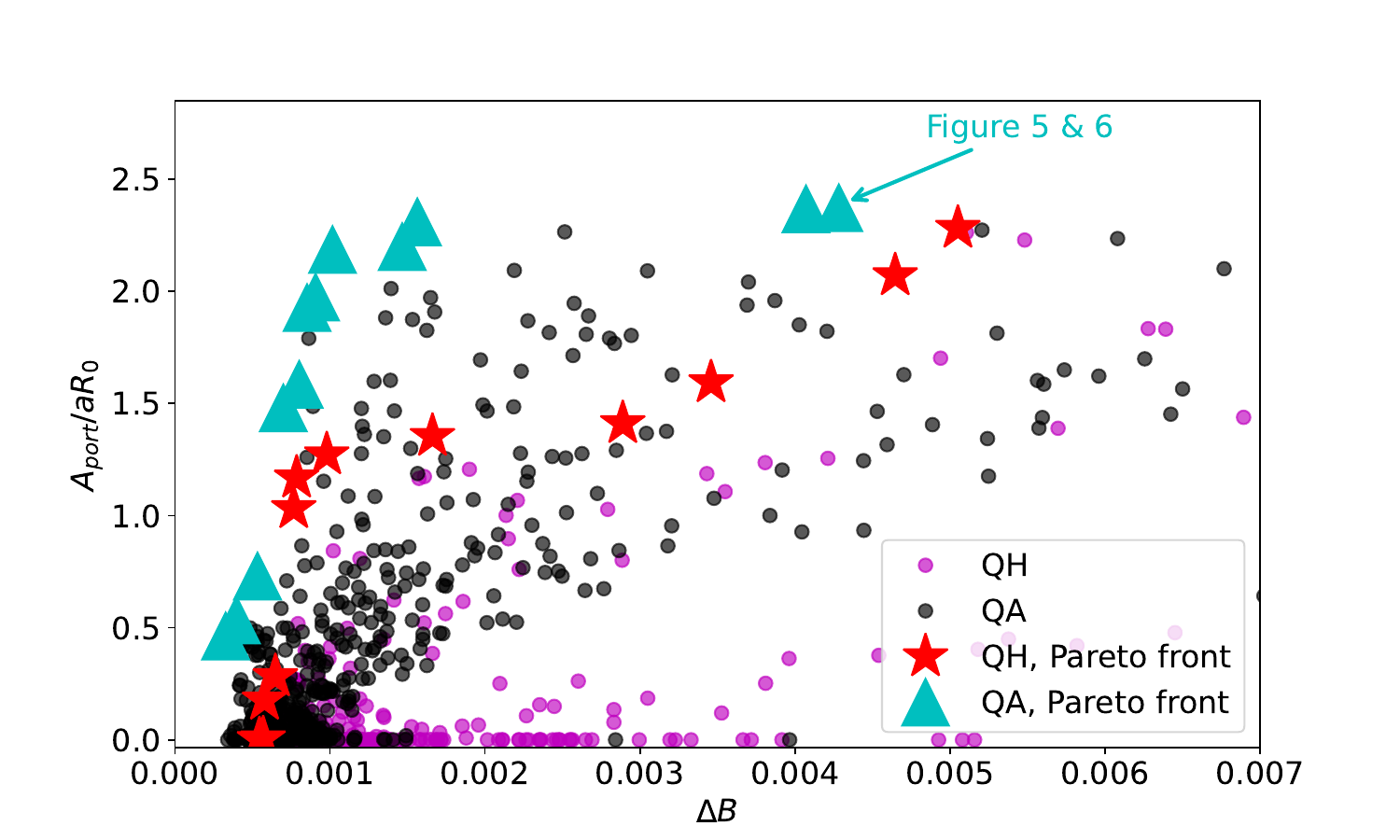}
    \caption{Port size for horizontal access (top) and vertical access (bottom) as a function of the averaged field error for multiple minima. Each point represent an extremum found by the minimizer for a specific initial guess and value of $w_{\text{area}}$. The Pareto front is highlighted with cyan triangles and red stars for the precise QA and precise QH configurations, respectively. Notice the difference in scale between the top and bottom panels. The examples shown in Figure \ref{fig:stage_II_radial_QH_3D_plot} - \ref{fig:stage_II_vertical_QA_field_quality} are highlighted with an arrow on both panels.}
    \label{fig.pareto}
\end{figure}

Figure \ref{fig.pareto} shows the Pareto front for a radial access port (top panel) and vertical access port (bottom panel) to the precise QA and precise QH configurations. First, we observe that radial ports saturates at a smaller size in the case of the precise QH in comparison to the precise QA configuration. In addition, many solutions found in the case of the precise QH configuration have very small ports, with an enclosed area close to zero. Regarding vertical access, the optimizer leads to similar results for the QA and QH configurations.

Overall, it seems that the optimization leads to better results in the case of the precise QA configuration. One possible explanation is that the precise QA configuration has a greater magnetic gradient scale length of $L_{\nabla B}=5.3$, in comparison to $L_{\nabla B}=3.54$ for the precise QH, which correlates with the required coil-plasma distance \citep{kappel_2024a}. This indicates that some plasma boundaries are more suitable than others to obtain large access ports, and a single-stage optimization algorithm could be used to find optimized plasma and coils that accept large access ports. This line of research is left for future work.

\begin{figure}
    \centering
    \includegraphics[width=0.75\linewidth]{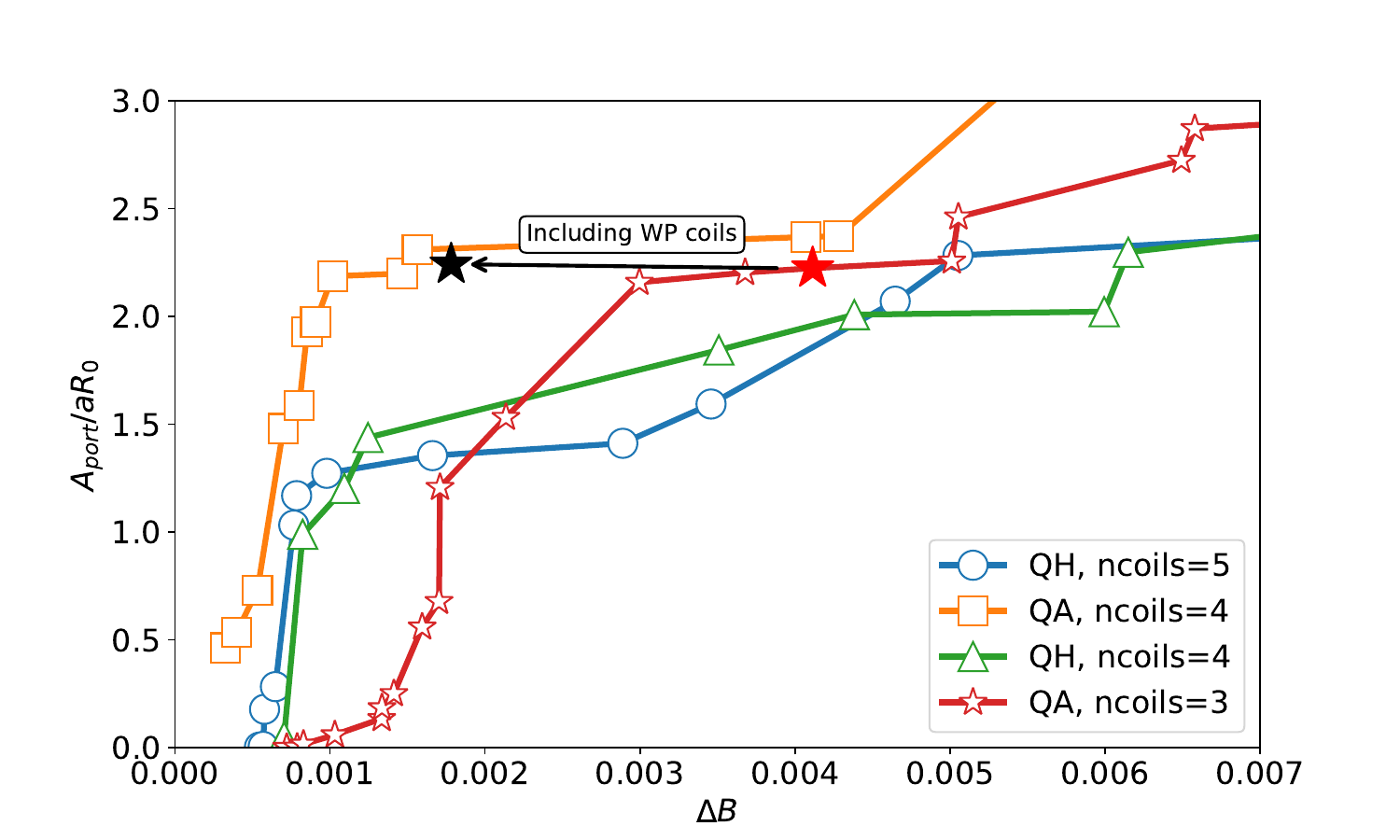}
    \caption{Pareto front formed by the normalized vertical access port size and the averaged normalized field error. Comparison between the precise QH with 4 and 5 coils for field periods, and the precise QA with 3 and 4 coils per field periods. The red star shows the highlights the configuration without windowpane coils discussed in section \ref{sec:wps}, while the black star shows the same configuration when windowpane coils are used to reduce the quadratic flux. The arrow underscores the improvement in field quality obtained by including windowpane coils.}
    \label{fig:pareto_ncoils}
\end{figure}

The maximum achievable access port size depends on the number of modular coils; intuitively, one expects that larger ports could be achievable with less coils at the cost of larger normalized field error. This trade-off is explored by looking at the Pareto front for different number of coils per field period (see Figure \ref{fig:pareto_ncoils}). While removing one modular coil from the precise QA configuration impacts negatively its Pareto front, \textit{i.e.} a larger field error is achieved for a similar port size, the precise QH configuration shows a similar Pareto front if 4 or 5 coils are considered per field period. With these Pareto fronts, it is then possible to determine the optimal number of coils given a target plasma boundary, maximum averaged field error tolerable, and target port size. For example, for the precise QA configuration, if a field error below $0.5\%$ is targeted, and a port of size $A_{\text{port}}/aR_0\geq 1.0$ is desired, a design with four coils per field period is required. If a field error of $1\%$ is acceptable, then a configuration with three coils per field period is possible. Of course, this choice then depends on other physics and engineering considerations.

\begin{figure}
    \centering
    \hfill
    \includegraphics[width=0.4925\linewidth]{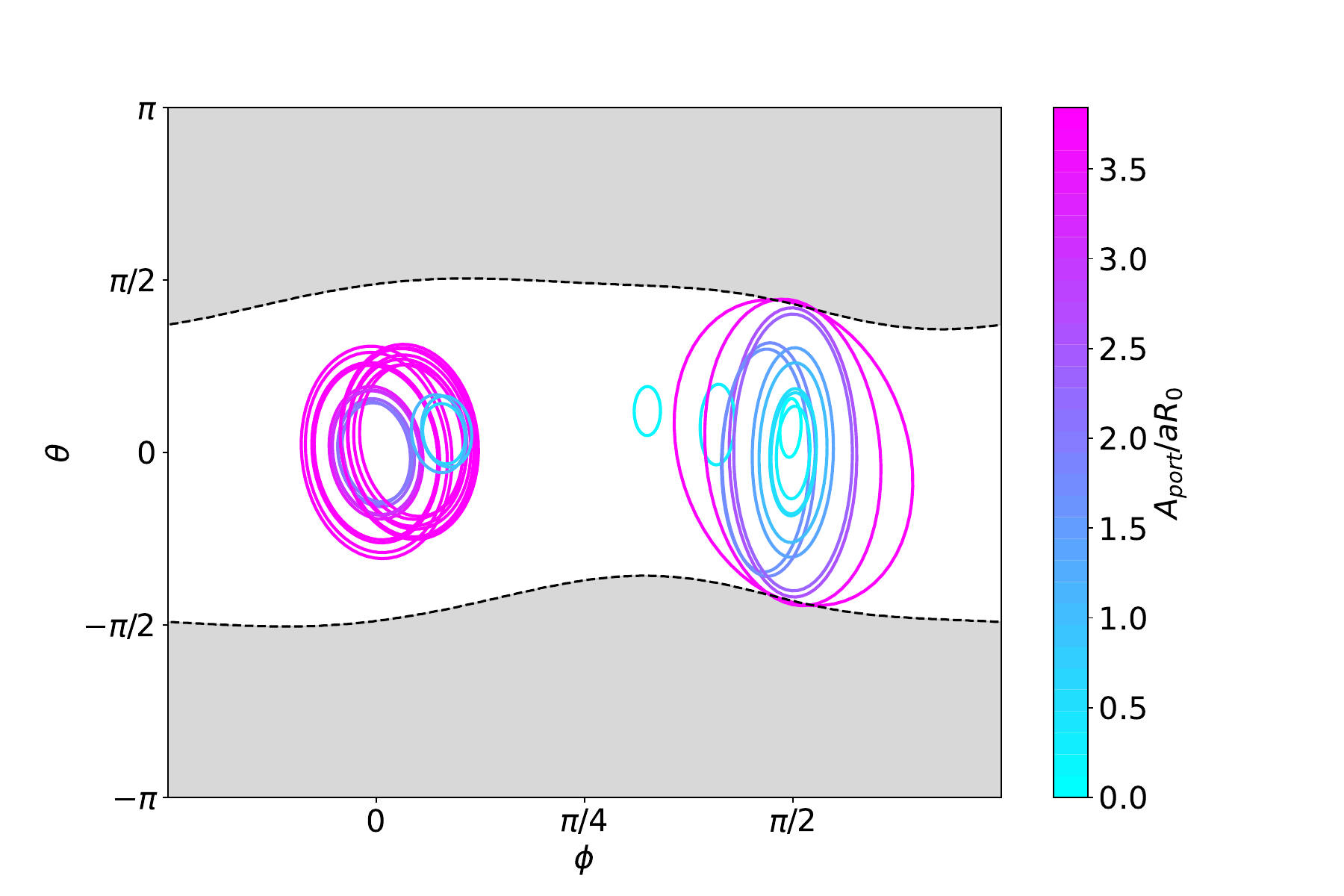}
    \hfill
    \includegraphics[width=0.495\linewidth]{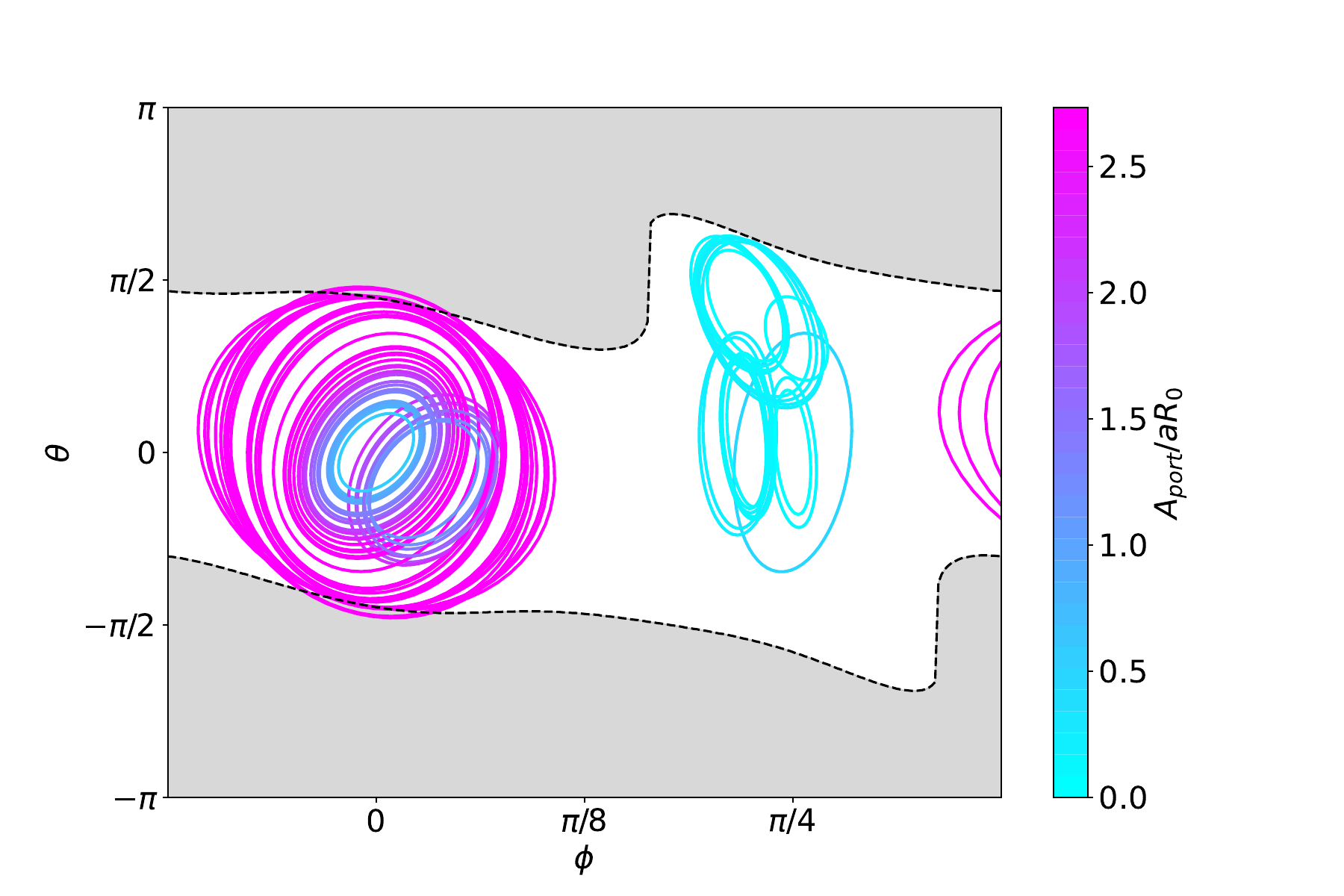}
    \hfill 
    \\
    \hfill
    \includegraphics[width=0.495\linewidth]{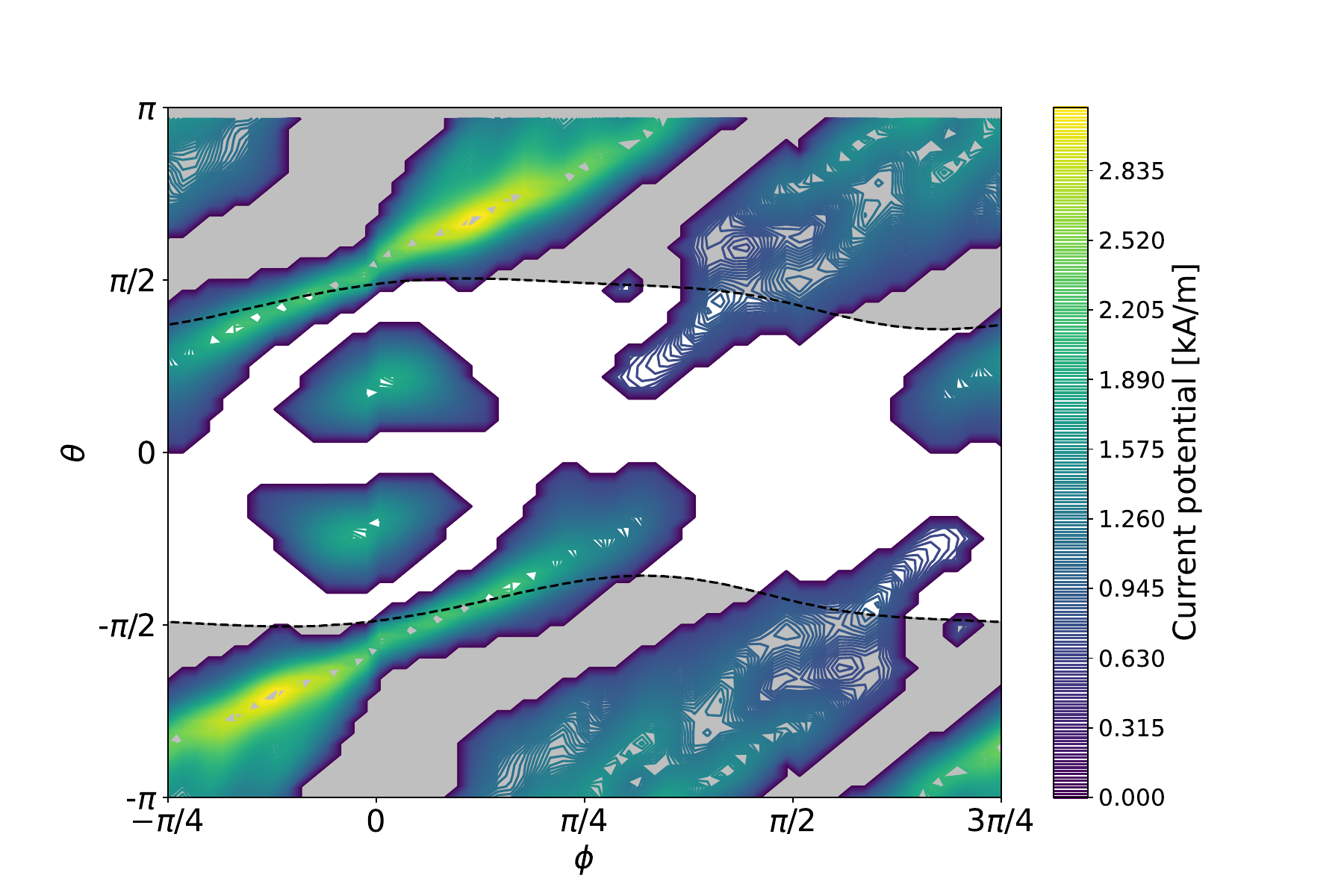}
    \hfill
    \includegraphics[width=0.495\linewidth]{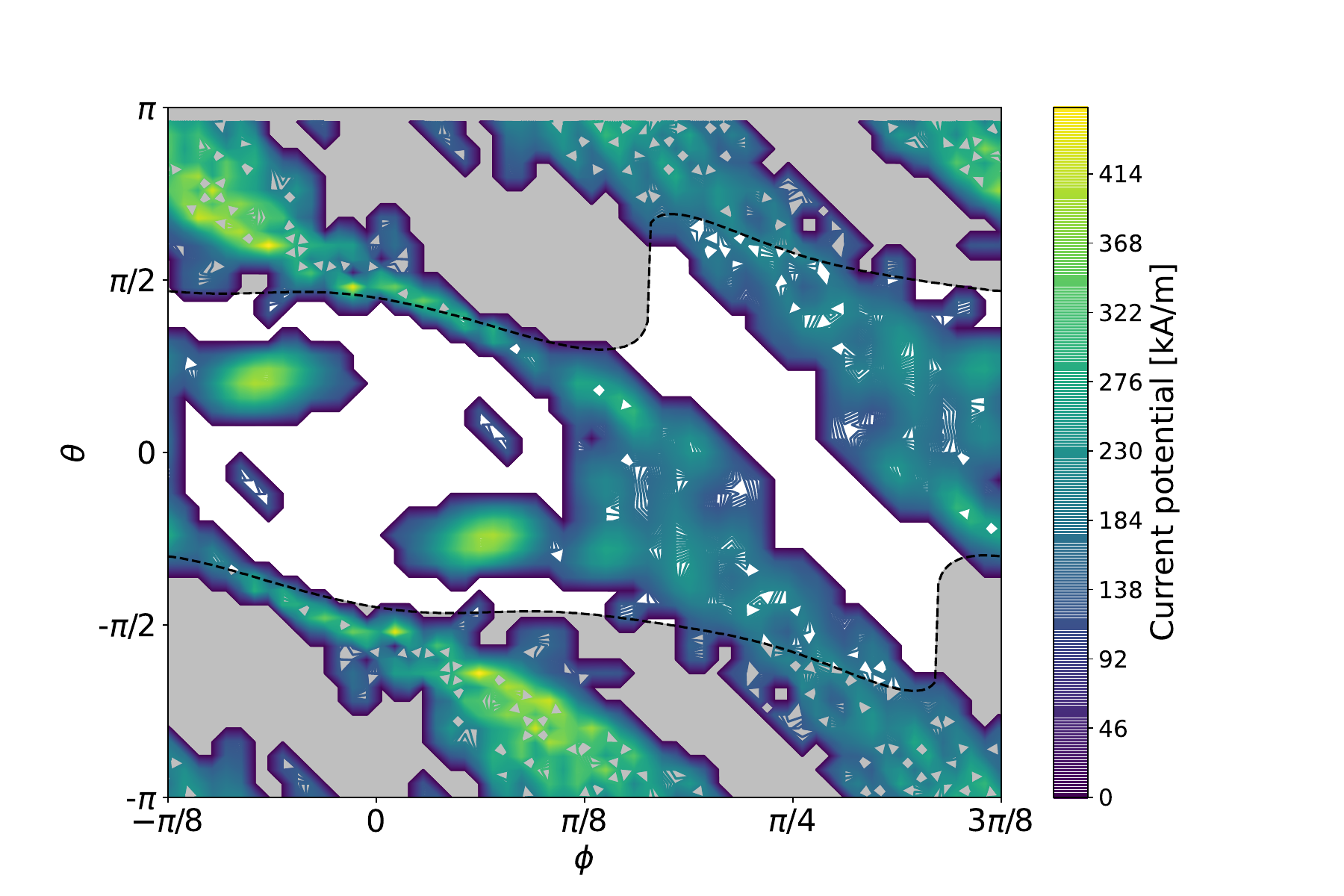}
    \hfill
    \caption{Top: Optimized radial access port location for the precise QA configuration (left) and precise QH configuration (right). Curve colors indicate the normalized port area. The grayed area indicates the region where the forward-facing port penalty, Eq.(\ref{eq.ffp}), is not satisfied. Bottom: \blue{Current potential patches showing radial port locations with 50\% of currents on the winding surface removed, shown in white. The regions of inefficient current placement, in white, correlate strongly to the locations of the radial ports found shown on the top panels. The radial ports coincided with transfer matrices (Eq. (\ref{eq.transfer_matrix})) of lower condition numbers, suggesting the ease of placement of radial ports for these configurations.}}
    \label{fig:RA_port_location_distribution}
\end{figure}

Interestingly, most configurations find that the largest achievable port is located at one specific location on the plasma boundary. On Figure \ref{fig:RA_port_location_distribution}, the final port locations are shown for all configurations shown on Figure \ref{fig.pareto} satisfying $\max(\mathbf{B}\cdot\mathbf{n}/B)\leq 0.5\%$ and $A_{\text{port}}/aR_0 \geq 0.025$. The final distribution of ports is showing an accumulation of ports at two specific points, either at $\phi_0=0$ or $\phi_0=\pi/N_{f}$, \textit{i.e.} at the beginning and center of a field period. This position is generally where the plasma surface bends, \textit{i.e.} where the plasma surface has large curvature. The largest ports found are located at $\phi=0$, while smaller ports are found around $\phi=\pi/N_f$.

\begin{figure}
    \centering
    \hfill
    \includegraphics[width=0.495\linewidth]{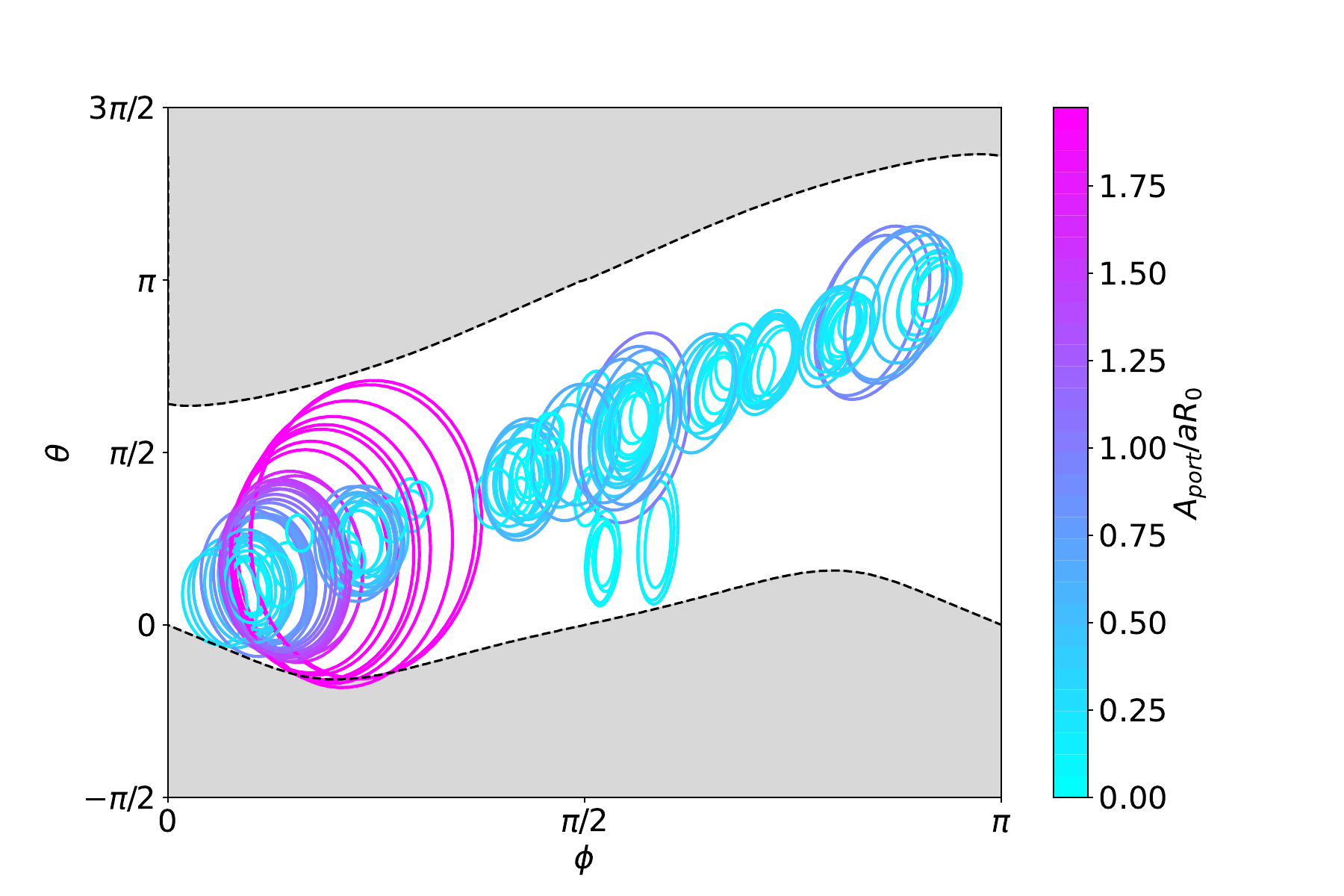}
    \hfill
    \includegraphics[width=0.495\linewidth]{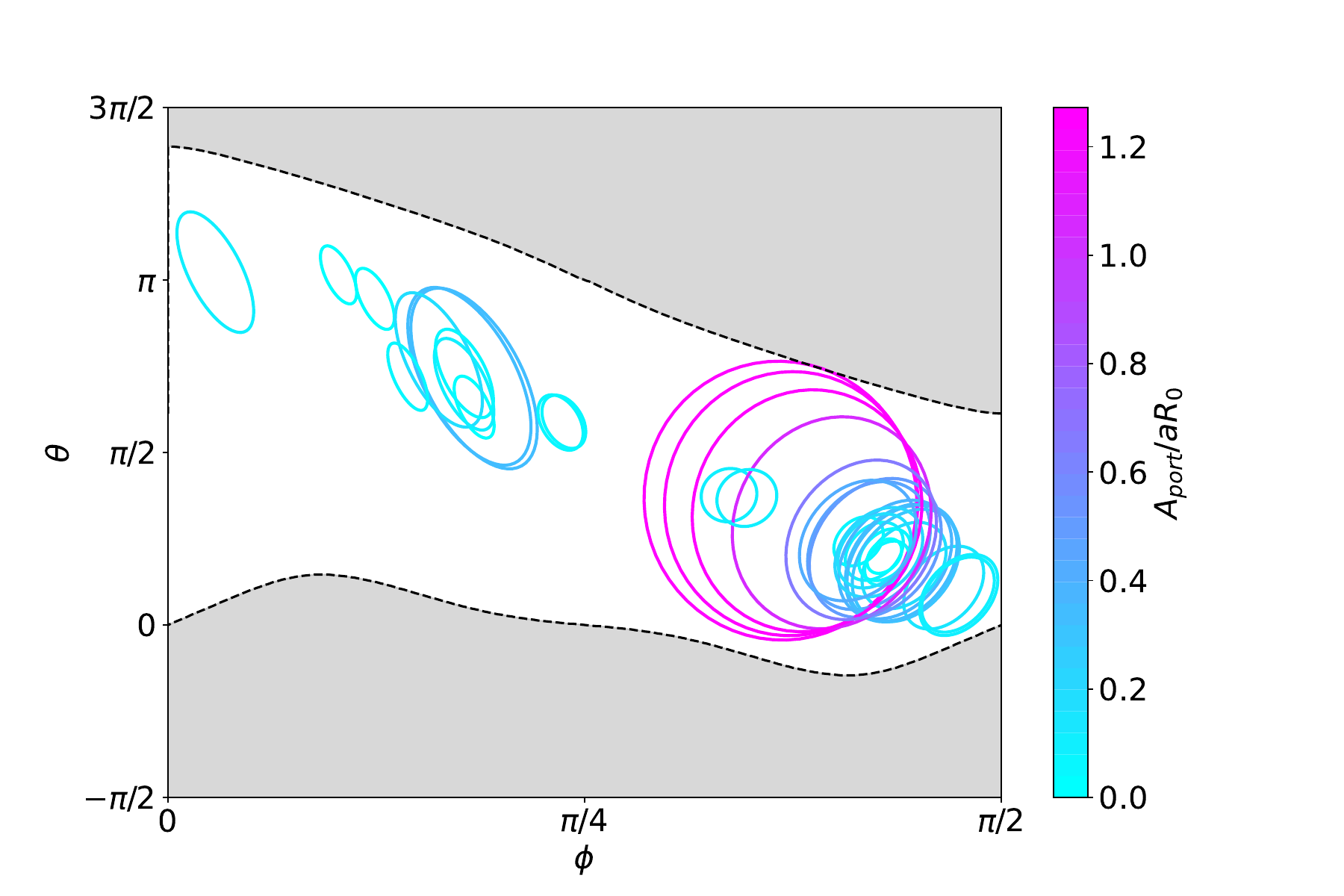}
    \hfill
    \\
    \hfill
    \includegraphics[width=0.495\linewidth]{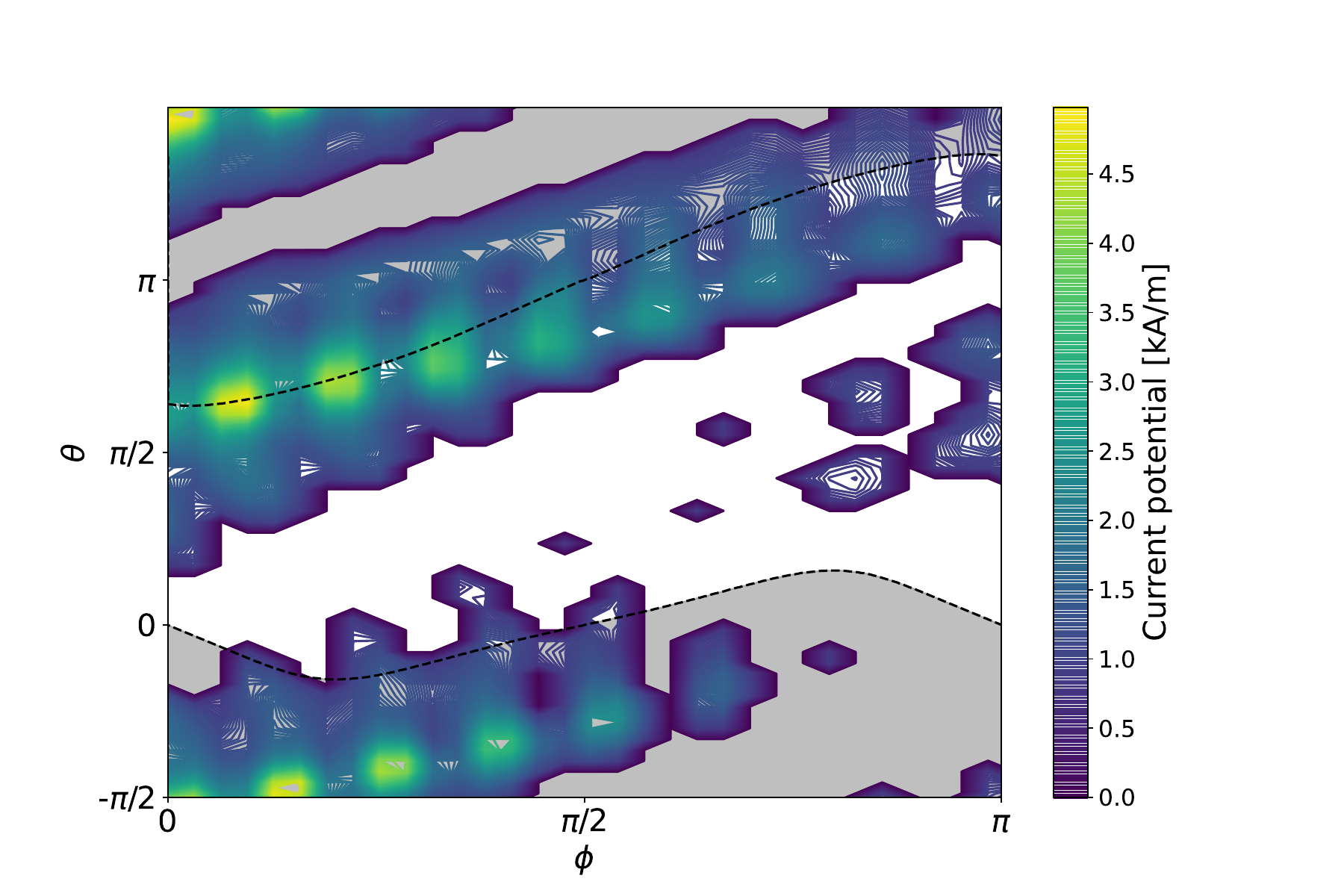}
    \hfill
    \includegraphics[width=0.495\linewidth]{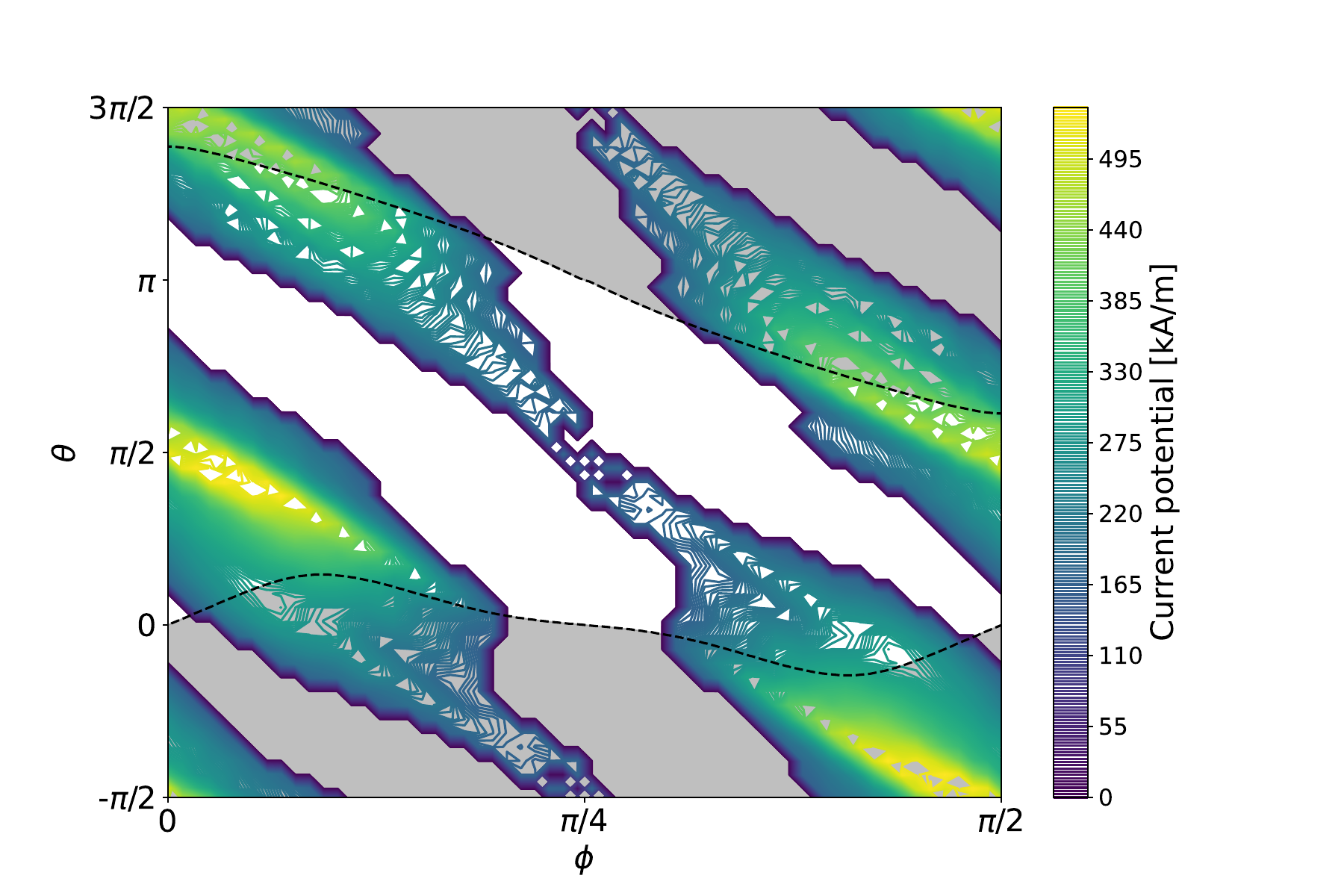}
    \hfill
    \caption{Top: Optimized vertical access port location for the precise QA configuration (left) and precise QH configuration (right). Curve colors indicate the normalized port area. The grayed area indicates the region where the forward-facing port penalty, Eq.(\ref{eq.ffp}), is not satisfied. Bottom: \blue{Current potential patches showing vertical port locations with 50\% of currents on the winding surface removed, shown in white. The regions of inefficient current placement, in white, correlate strongly to the port locations shown on the top panel. Interestingly, the vertical ports coincided with transfer matrices (Eq. (\ref{eq.transfer_matrix})) of higher condition numbers, suggesting the placement of vertical ports to be more difficult for these configurations.}}
    \label{fig:VA_port_location_distribution}
\end{figure}

A similar comparison can be done for vertical access ports. Figure \ref{fig:VA_port_location_distribution} shows the toroidal distribution of vertical access ports for the precise QH and precise QA configurations over a full field period. Here the full field period is shown, as upward vertical access ports on the first half field period correspond to vertical access ports pointing downward on the second half field period; to explore all upward vertical facing ports the full field period has to be considered. Again, we observe a preferred location for ports, although the result is less striking than in the radial access case. This result clearly indicates that some geometrical properties of the plasma boundary can be beneficial to large port access. Identifying a metric that relates the plasma shape to port sizes could help find plasma boundaries more suitable for both horizontal and vertical access.

\blue{The locations of the resulting vertical and radial ports may be understood using a current potential approach known as current potential patches (CPPs) \citep{elder_2024}. The strength of each individual dipole, represented in an array $\mathbf{d}_\text{dipole}$, is obtained my minimizing
\begin{align}
    f_\text{dipole} = |\mathbf{A}_\text{dipole}\cdot \mathbf{d}_{dipole} - \mathbf{B}_\text{coils}\cdot\hat{n}|^2, \label{eq.transfer_matrix}
\end{align}
where $\mathbf{B}_\text{coils}$ is the magnetic field produced by the coils, and $\mathbf{A}_\text{dipole}$ is the transfer matrix of the dipole magnetic field. The condition number of $\mathbf{A}_\text{dipole}$ provides information about the sensitivity of the dipole field to perturbation, and can be understood as an indicator for the difficulty to generate this field with realistic coils. More details about the CPPs method and the interpretation of the transfer matrix condition number can be found in \citep{elder_2024}.

For a current sheet flowing on winding surface $W$ to produce a magnetic surface $\Gamma$, some locations of current flow on $W$ are less efficient at minimizing the normal field error on $\Gamma$ than others.} 
The CPPs are shown on the lower panels of Figure \ref{fig:RA_port_location_distribution} and \ref{fig:VA_port_location_distribution}. Here, the least efficient CPPs were removed until only $50\%$ of the winding surface remained covered by CPPs, effectively generating ``holes" in the CPPs locations. Note that reducing the number of CPP on the winding surface generally increases the condition number of the transfer matrix as defined in \ref{eq.transfer_matrix}. Interestingly, the location of these holes seems to correlate with the optimum port location shown on the top panels of Figure   \ref{fig:RA_port_location_distribution} and \ref{fig:VA_port_location_distribution}. Even though the comparison is less striking for vertical access ports, the method of CPPs gives useful information regarding the optimum port location on a plasma boundary. One possible pathway to improve stellarator accessibility could therefore be to target plasma shapes where their large sections of their corresponding winding surface remain free of CPPs. \blue{Note that the CPPs method is sensitive to the choice of winding surface. Here the winding surfaces that support the Wiedman \citep{wiedman_2023} and Wechsung \citep{wechsung_2022b} coil sets were considered as well as conformal surfaces slightly offset from the plasma surface. It was found that the CPPs computed using the conformal winding surfaces correlated better to the port access results.}

\section{Including windowpane coils}\label{sec:wps}
As observed in the previous optimizations, increasing the port size in general increases the value of $J_{\text{quadflux}}$, \textit{i.e.} decreases the capability of the coils to generate the target plasma boundary. One possible way to compensate this error is to include additional magnetic field sources that do not get in the way of an access port. These sources can be, for example, windowpane coils, as in the Eos design \citep{kruger_2025,gates_2025b}, permanent magnets \citep{helander_2019,qian_2022,kaptanoglu_2022,zhu_2022,qian_2023}, or passive superconducting conductors \citep{kaptanoglu_2025}. We now show that a similar set-up as what has been proposed by \citet{gates_2025b} can improve a configuration beyond the Pareto front shown on Figure \ref{fig:pareto_ncoils}. We focus on the precise QA configuration with three coils per half field period; as seen on Figure \ref{fig:pareto_ncoils}, the field error is strongly impacted when three coils per field period are considered instead of four. A configuration with less coils is however beneficial to reduce the overall device complexity. To get the best of both worlds, we show here that a satisfactory field error, and a large vertical access port can be obtained with only three coils per half field period when the plasma is partially shaped by windowpane coils. In this section, the plasma boundary, coils, and currents are scaled to match the ARIES-CS volume ($V=444\ \text{m}^3$) \citep{ku_2008} and magnetic field on axis ($B=5.7$ T) \citep{najmabadi_2008}.

We construct an array of windowpane coils surrounding the plasma, and optimize their currents. We describe windowpane coils as planar curves, shaped as rectangles with rounded corners, facing the plasma boundary. The center of each windowpane coil lies on a winding surface, constructed as the set of points one meter away from the plasma boundary. The construction of the windowpane array uses a library that will be described in a future publication \citep{Halpern_Windowpane_Coil_Optimization_2025}, and relies on multiple parameters, such as the windowpane grid resolution, the shaping of windowpane coils, and the spacing between each windowpane coil. Here a set of parameters that leads to reasonable results is chosen.

The optimization of the modular coils, windowpane coils, and vertical access ports is then performed in a stepped approach. First, the modular coils and the port are optimized, similarly to what was described in section \ref{sec:stage_II}. Then, the modular coils and the port are fixed, and the current in the windowpane coils is optimized to minimize the quadratic flux on the plasma boundary. Finally, windowpanes with a small current relative to the other coils (\textit{i.e.} smaller than $200$ kA) are deleted to promote sparsity. These three steps are repeated until the current in the windowpane coils converges. Convergence is usually observed in 10 to 20 iterations.

\begin{figure}
    \centering
    \begin{tikzpicture}
    \node (t1) at (-4,2) {\includegraphics[width=.475\linewidth]{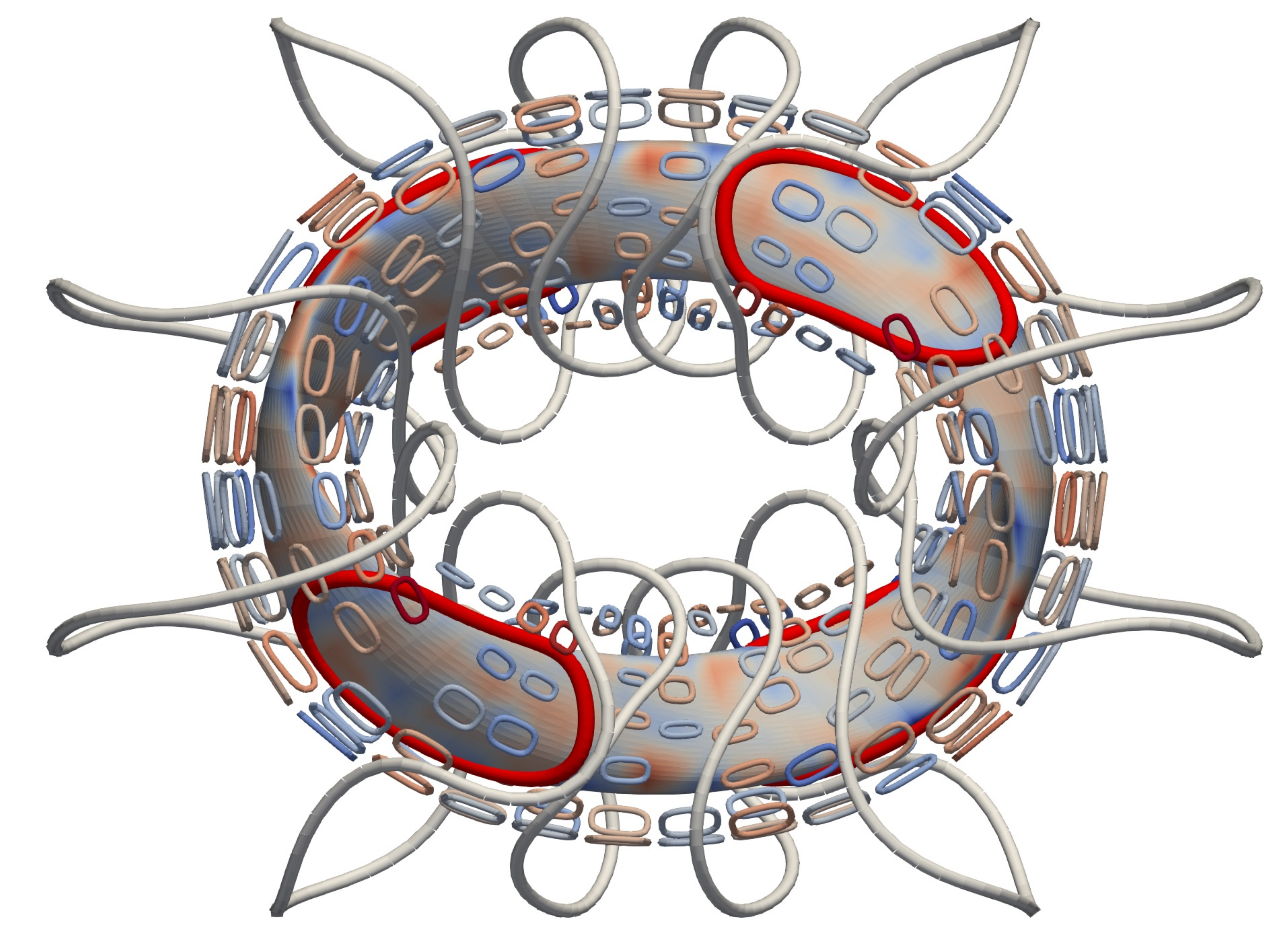}};
    \node (t3) at (4,2) {\includegraphics[width=.475\linewidth]{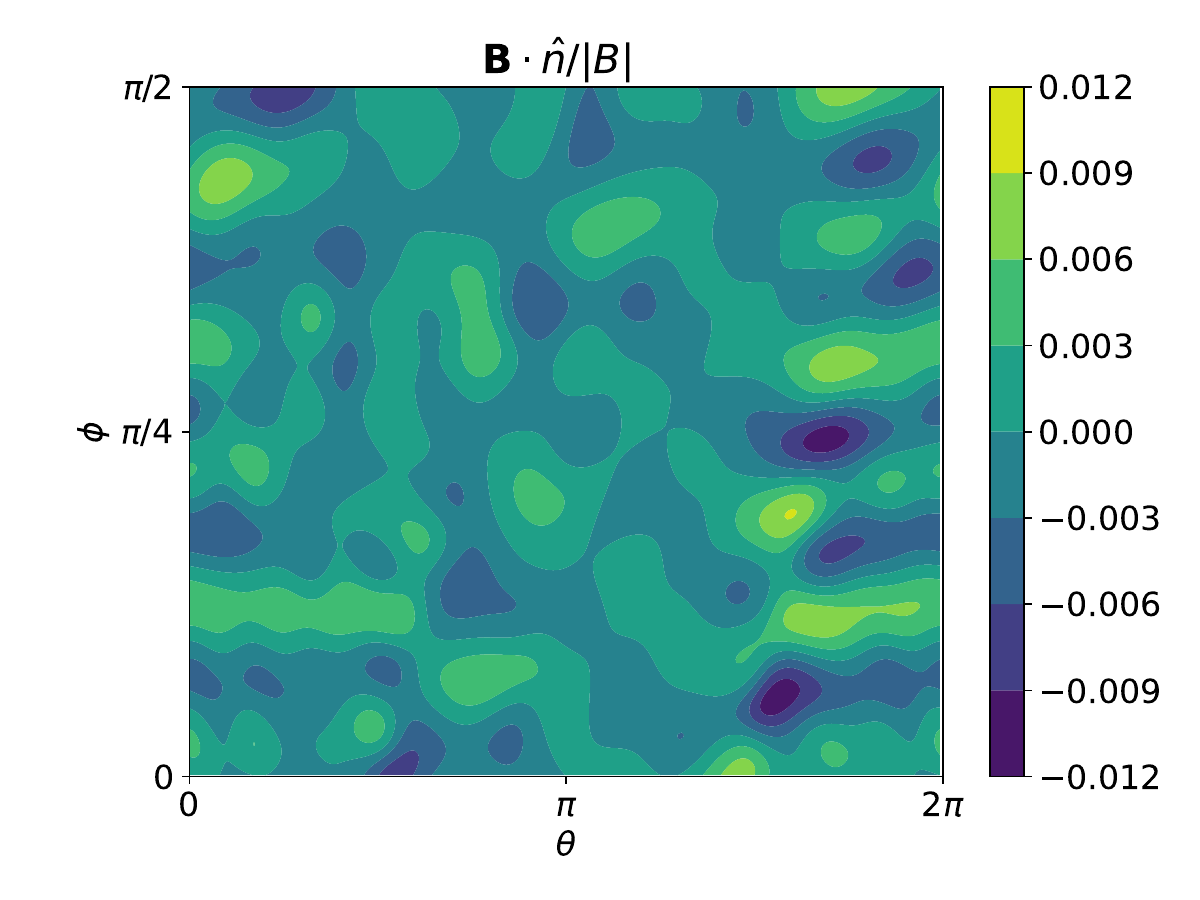}};
    \node (t2) at (0,-6) {\includegraphics[width=.8\linewidth]{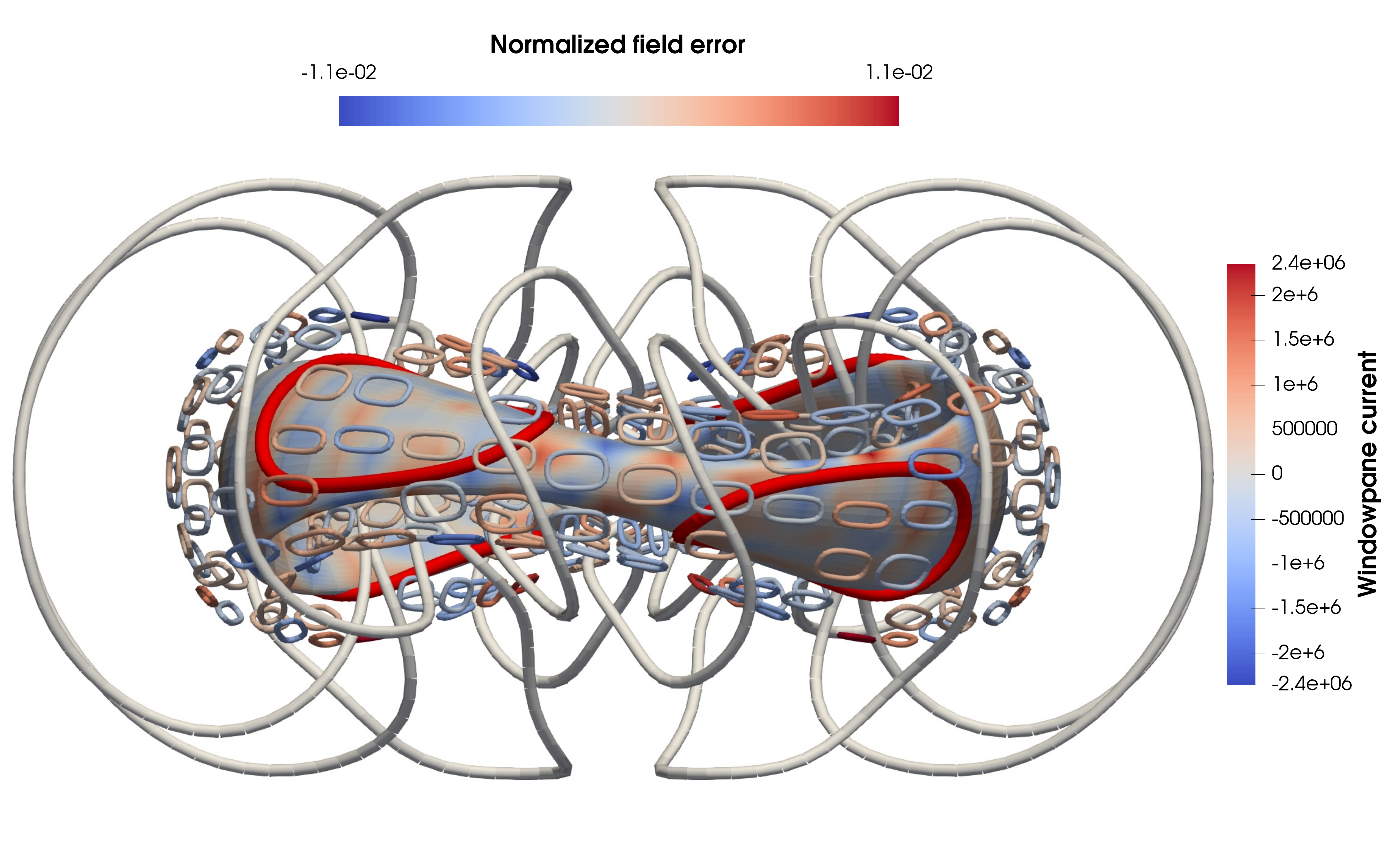}};
    \end{tikzpicture}
    \caption{Top view (top left) and side view (bottom) of the precise QA configuration with vertical access port and windowpane coils. Modular coils are plotted in gray, windowpane coils in blue, ports in red, and the colors on the plasma boundary are the normalized field error, \textit{i.e.} $\mathbf{B}\cdot\mathbf{n}/B$. Top right: Normalized field error over one half field period.}
    \label{fig:wp_method1_3d}
\end{figure}

\begin{figure}
    \centering
    \begin{tikzpicture}
        \node (t1) at (-4,6) {\includegraphics[height=5cm]{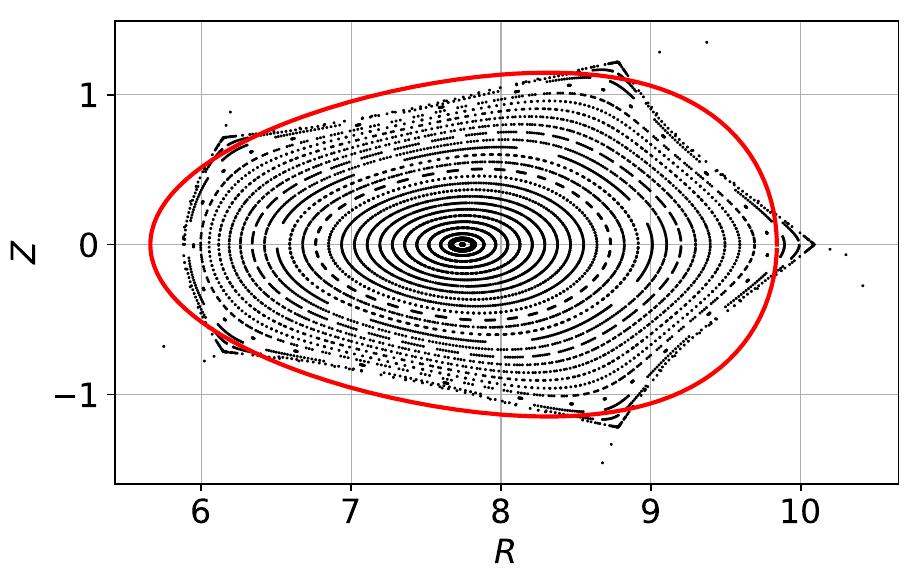}};
        \node (t4) at ( 5.5,0.5) {\includegraphics[height=7.5cm]{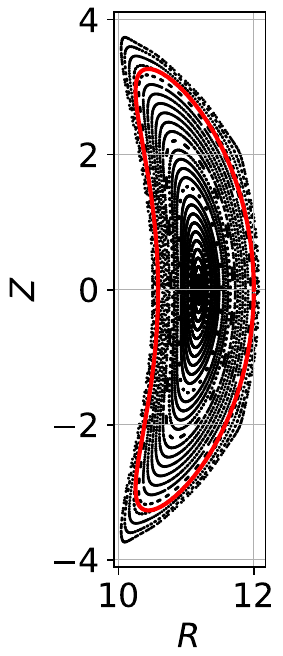}};
        \node (t2) at ( 2.5,5.5) {\includegraphics[height=7.5cm]{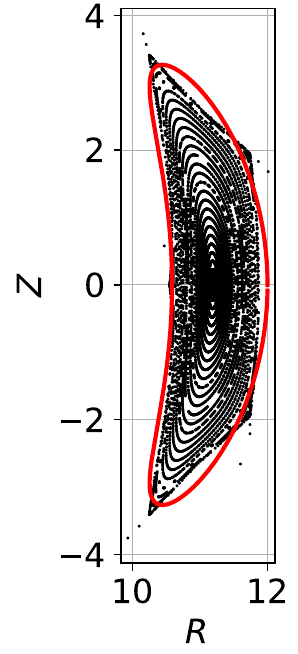}};
        \node (t3) at (-4,0) {\includegraphics[height=5cm]{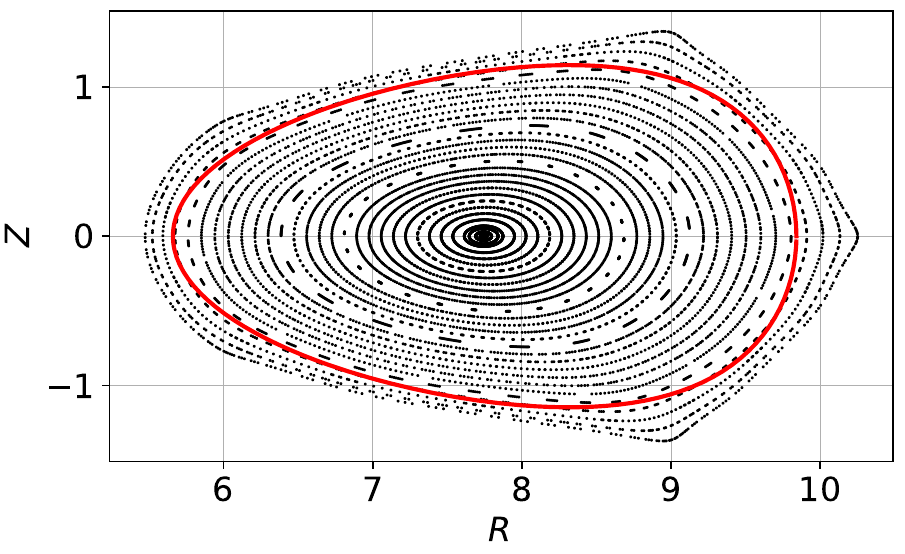}};
    \end{tikzpicture}
    \caption{Poincar\'e section (in black) at the triangular cross-section (left) and bean cross-section (right) for the precise QA configuration with $N_{\text{coils}}=3$ and where a vertical access port is targeted. Top: no windowpane coils, bottom: with windowpane coils. The red curve shows the target plasma boundary.}
    \label{fig:wp_method1_poincare}
\end{figure}

As an example, we optimize a set of windowpane coils to correct the field error on the precise QA plasma boundary, using $N_{\text{coils}}=3$ and targeting a large vertical access port.  If no windowpane coils are considered, a configuration with an access port of normal size $A_{\text{port}}/aR_0=2.22$ is obtained, with an averaged field error of $\Delta B=4.11\cdot 10^{-3}$. As a consequence, the plasma boundary does not match with the targeted surface (see top panels of Figure \ref{fig:wp_method1_poincare}). Including windowpane coils slightly increases the port size to $A_{\text{port}}/aR_0=2.24$, and decreases the averaged field error to $\Delta B=1.78\cdot 10^{-3}$, a decrease of about $57\%$! The optimized coil set, as well as the Poincar\'e section of the magnetic field, are shown on Figure \ref{fig:wp_method1_3d} and \ref{fig:wp_method1_poincare}. The target plasma boundary is qualitatively better recovered (see lower panels of Figure \ref{fig:wp_method1_poincare}). Overall, this configuration with windowpane coils is better than any other configuration on the Pareto front shown on Figure \ref{fig:pareto_ncoils} (black star). The currents required in the windowpane coils are all below $2.4$ MA, which is of the same order as the maximum current considered for the Eos design \citep{kruger_2025}. The example shown here is, however, an extreme case; smaller ports, as shown on Figure \ref{fig.pareto}, generate lower field errors, and thus require smaller correction from windowpane coils, \textit{i.e.} smaller currents. Further studies would be required to explore the Pareto front formed by the port size and quadratic flux objectives when windowpane coils are used; the workflow used here is however not applicable to this use case, as it requires significant human inputs to initialize a reasonable windowpane array.

\section{Conclusion}\label{sec:conclusion}
The design of future fusion power plants must incorporate an efficient maintenance strategy. In this work, we have introduced a novel optimization approach to design stellarator coils that allow for large access ports. The access port is represented as a closed curve on the plasma boundary, with its enclosed area maximized under constraints that ensure no coils obstruct the port. Particular attention has been given to formulating and implementing new objectives and constraints related to port optimization. 

This optimization framework was applied to different scenarios. First, we demonstrated that the largest possible port between two fixed coils can be found by optimizing only the port geometry. The approach was then extended to include simultaneous optimization of both the port and coils. It was observed that the coils deform and elongate to accommodate larger access ports. However, this study did not assess the feasibility of these coils from an engineering perspective, as such feasibility depends on the chosen conductor technology.

A trade-off between field quality and port size was explored through the Pareto front of the two competing objectives. It was found that relatively large radial access ports can be achieved without significantly degrading field quality, whereas even small vertical access ports require coil deformations that can severely disrupt magnetic surfaces. Additionally, we identified that large radial ports are naturally located in regions of high plasma boundary curvature, suggesting that introducing torsion in the plasma shape could facilitate the formation of vertical access ports. Finally, we demonstrated that adding windowpane coils can enhance the optimization results, yielding solutions superior to those on the Pareto front. Since windowpane coils can be removed during maintenance, they do not obstruct access paths. While windowpane coils were used as an illustrative example, alternative field-shaping techniques, such as permanent magnets \citep{helander_2019,qian_2022,kaptanoglu_2022,zhu_2022,qian_2023} or passive superconducting conductors \citep{kaptanoglu_2025}, could also be considered.

This port optimization framework opens several avenues for future research. First, single-stage optimization algorithms could be employed to explore configurations in which the plasma shape is directly modified to accommodate larger ports. This would enable a more integrated approach to balancing plasma performance with engineering constraints. Second, the method could be extended to allow for arbitrary access directions, rather than being limited to horizontal or vertical orientations. In principle, the access direction could be treated as an additional set of degrees of freedom, as suggested by the port locations in the shaded-out regions of Figures \ref{fig:RA_port_location_distribution} and \ref{fig:VA_port_location_distribution}. Finally, a dedicated metric for accessibility could be developed to guide plasma shape optimization. As observed in this study, large ports are naturally facilitated by certain plasma boundary features; explicitly favoring such configurations could further enhance port accessibility in stellarator design. 

While designing stellarators with large access ports is one approach to improving reactor accessibility, other strategies, such as sector-based maintenance, also hold significant potential. Optimizing stellarators for efficient sector-based maintenance is just as important as optimizing for port-based maintenance. A comparative study of these approaches would provide valuable insights into the trade-offs and feasibility of different maintenance strategies.

\section*{Acknowledgements}
We thank J. Lion, G. Stadler, A. Kaptanoglu, A. Giuliani for useful discussion. Special thanks to J. Halpern, for noticing that some coils resemble famous potato chips.  The authors acknowledge funding from Simons Foundation Targeted MPS Program (Award 1151685), the PPPL LDRD program, and the Simons Foundation MPS Collaboration Program (Award 560651). This research used resources of the National Energy Research Scientific Computing Center (NERSC), a Department of Energy Office of Science User Facility using NERSC award FES-ERCAP30322. T. Elder gratefully acknowledges funding from the Department of Energy Office of Fusion Energy Sciences Postdoctoral fellowship.

\appendix
\section{Radial access port, precise QA} \label{app:raqa}
\begin{figure}
    \centering
    \begin{tikzpicture}
    \node (t1) at (-4,2) {\includegraphics[width=.475\linewidth]{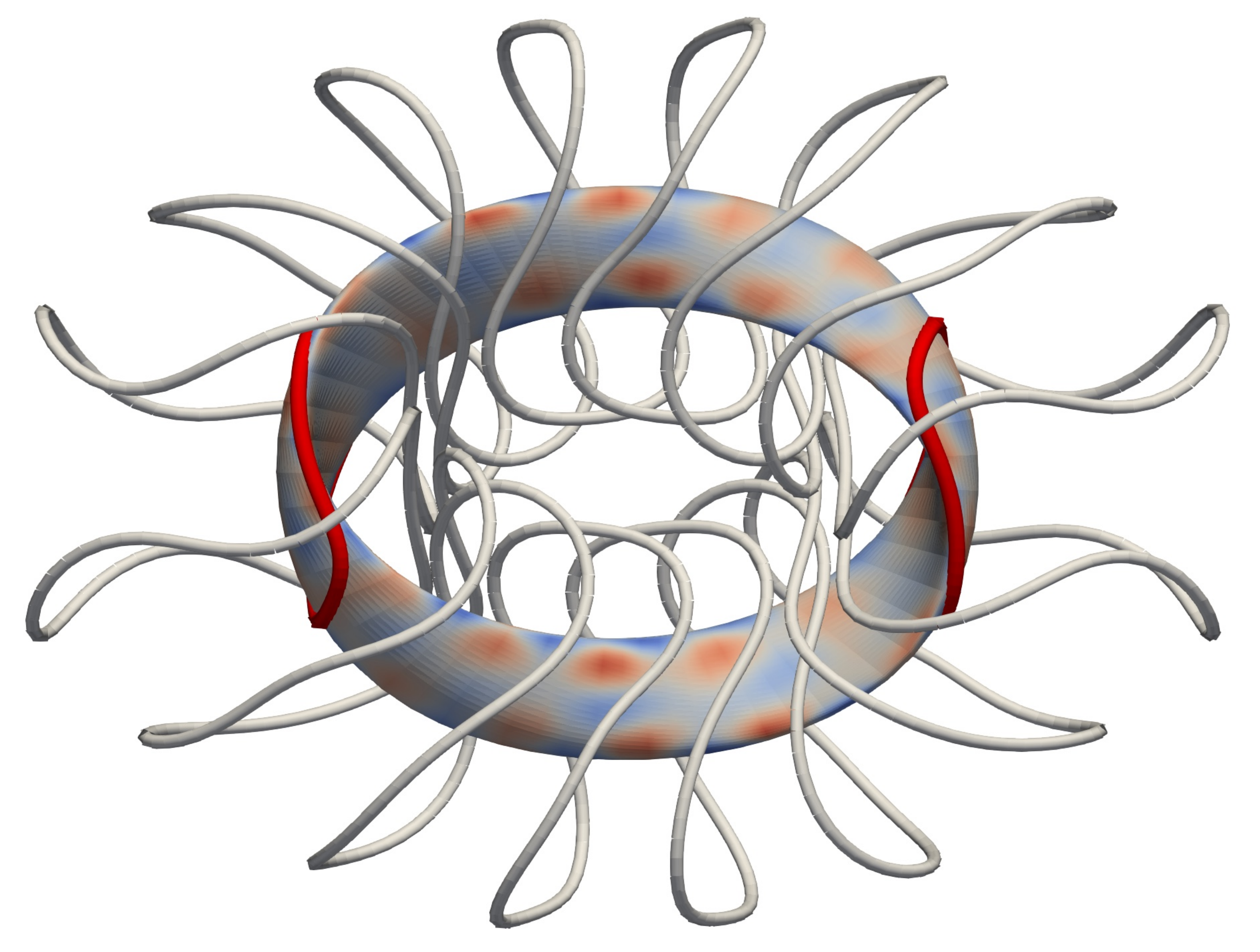}};
    \node (t3) at (4,2) {\includegraphics[width=.475\linewidth]{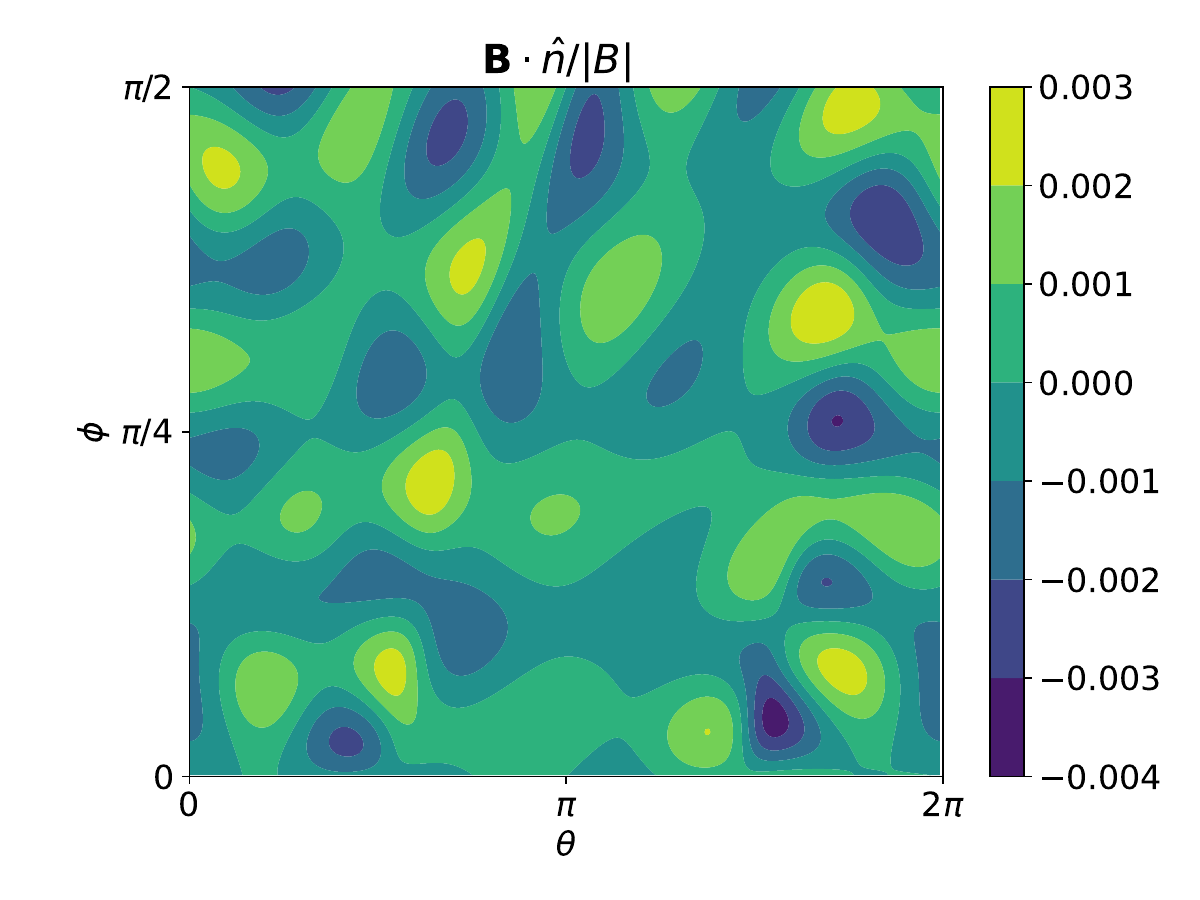}};
    \node (t2) at (0,-6) {\includegraphics[width=.8\linewidth]{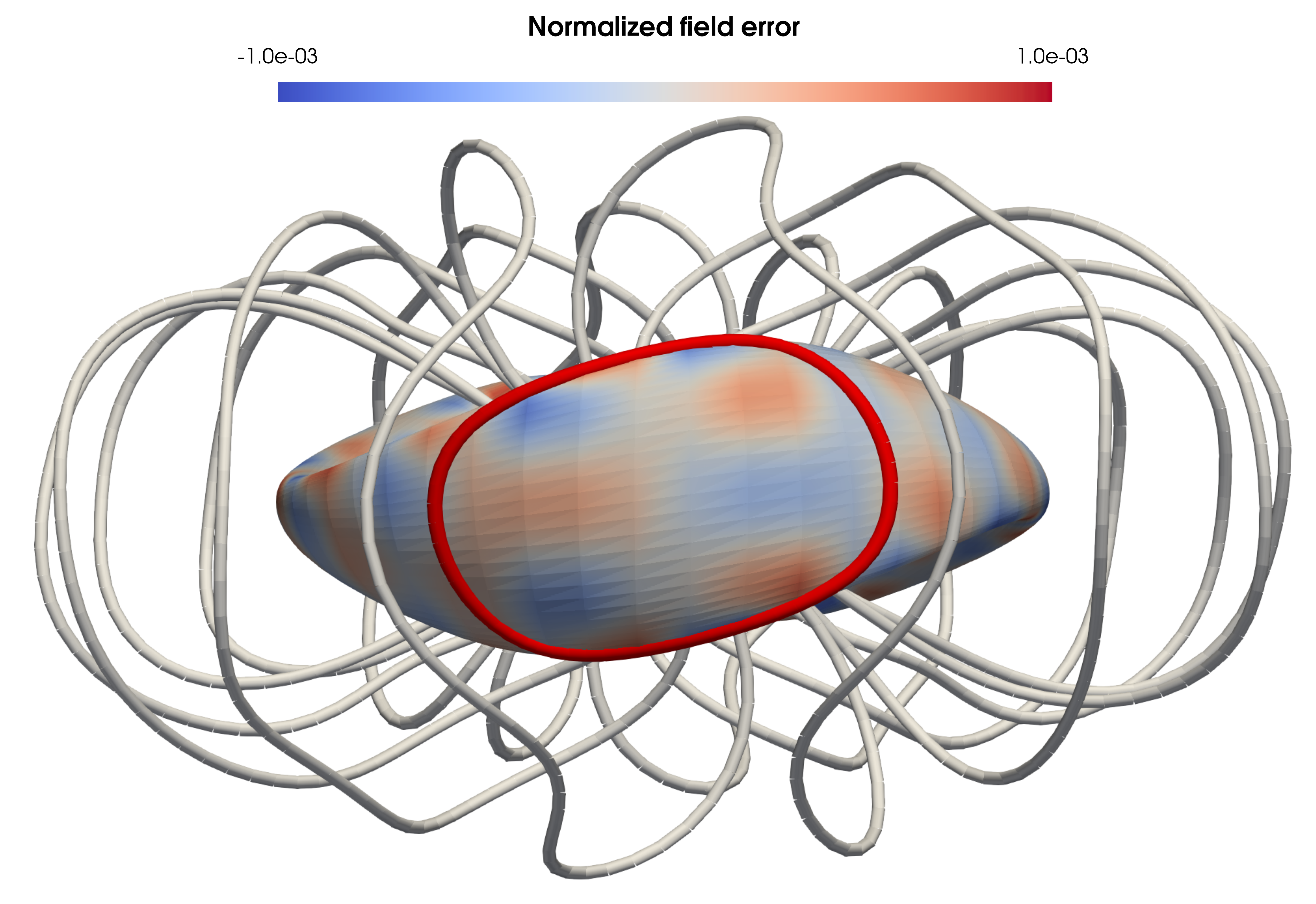}};
    \end{tikzpicture}
    \caption{Top view (top left) and side view (bottom) of the precise QA configuration with horizontal access port. Coils are plotted in gray, ports in red, and the colors on the plasma boundary are the normalized field error, \textit{i.e.} $\mathbf{B}\cdot\mathbf{n}/B$. Top right: Normalized field error over one half field period.}
    \label{fig:RAQA_example}
\end{figure}

\begin{figure}
    \centering
    \begin{tikzpicture}
        \node (t1) at (-4,0) {\includegraphics[height=5cm]{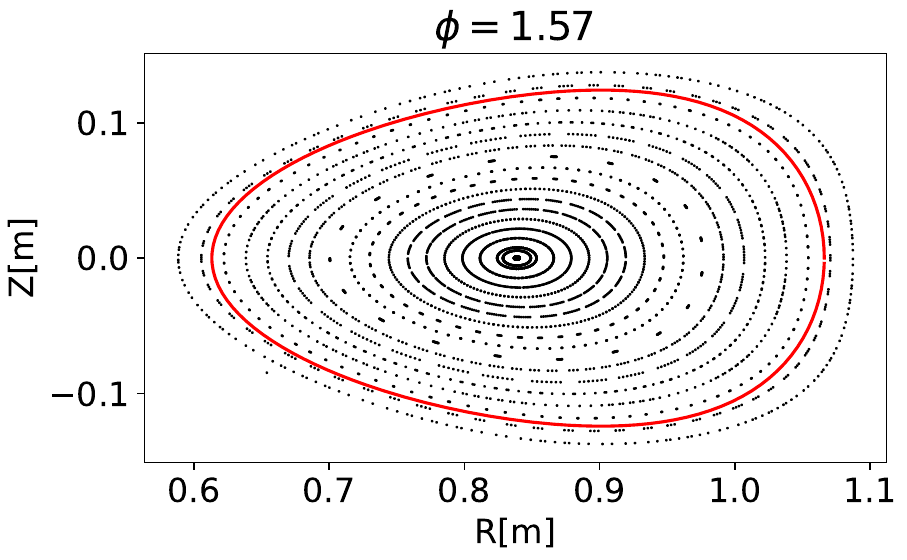}};
        \node (t2) at ( 4,0) {\includegraphics[height=7cm]{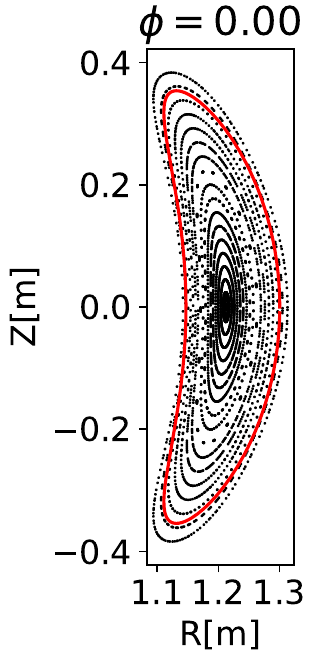}};
    \end{tikzpicture}
    \caption{Poincar\'e section (in black) at the triangular cross-section (left) and bean cross-section (right) for the precise QA configuration with radial access port. The red curve shows the target plasma boundary.}
    \label{fig:stage_II_RAQA_poincare}
\end{figure}

We provide an example of a stage II optimization in the case of the precise QA configuration with radial access ports. The normalized field error, as well as three-dimensional plots of the coils, the ports and the plasma boundary are shown on Figure \ref{fig:RAQA_example}, while Poincar\'e sections of the magnetic field are shown on Figure \ref{fig:stage_II_RAQA_poincare}. The obtained port has a normalized area of $A_\text{area}/aR_0 = 3.72$ for a maximum normalized field error of $\max\mathbf{B}\cdot\hat{n}/B = 0.3\%$. As in the case of the precise QH configuration with radial access ports, a relatively large port is obtained without compromising the field quality.

\section{Vertical access port, precise QH} \label{app:vaqh}
\begin{figure}
    \centering
    \begin{tikzpicture}
    \node (t1) at (-4,2) {\includegraphics[width=.475\linewidth]{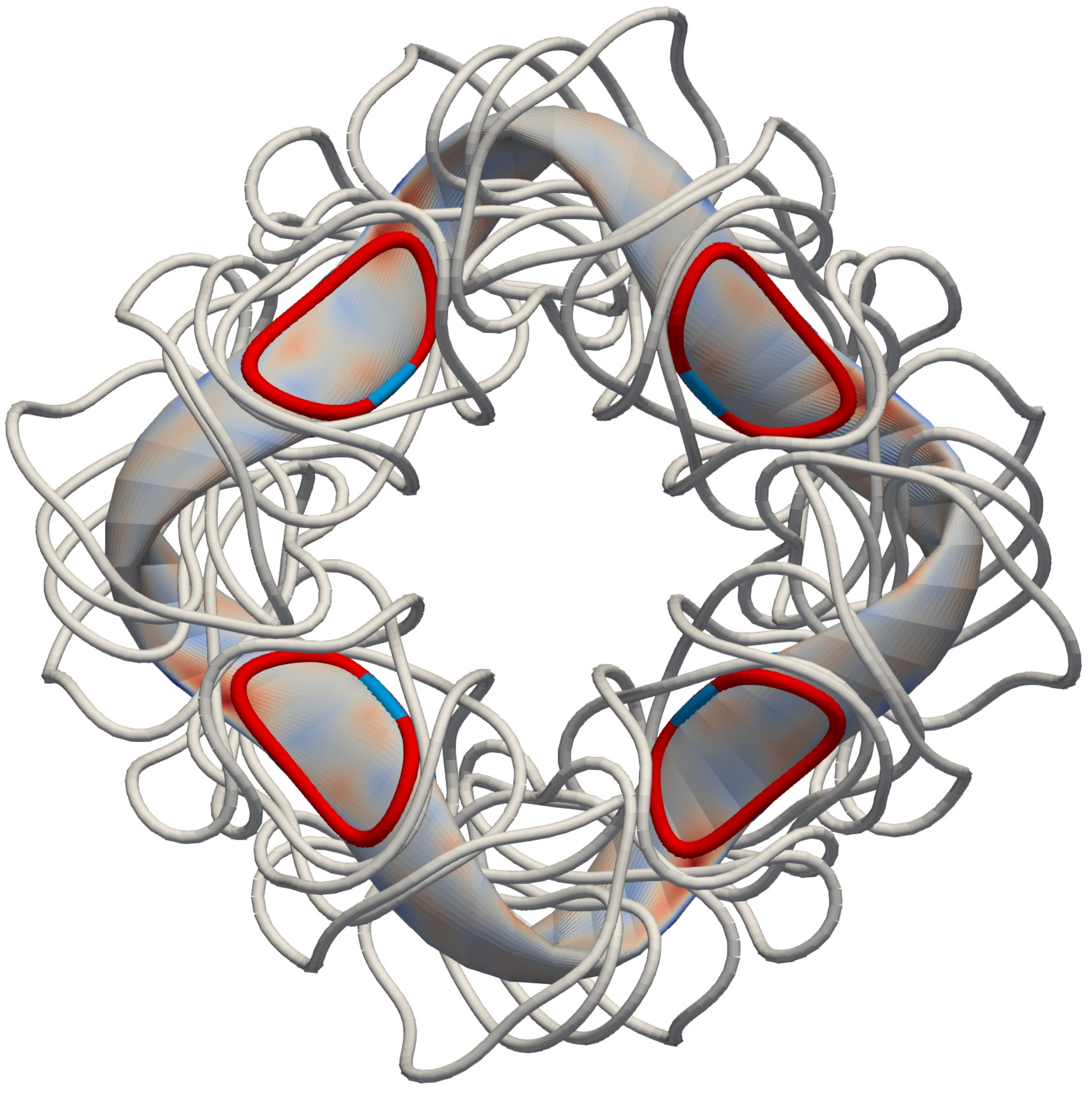}};
    \node (t3) at (4,2) {\includegraphics[width=.475\linewidth]{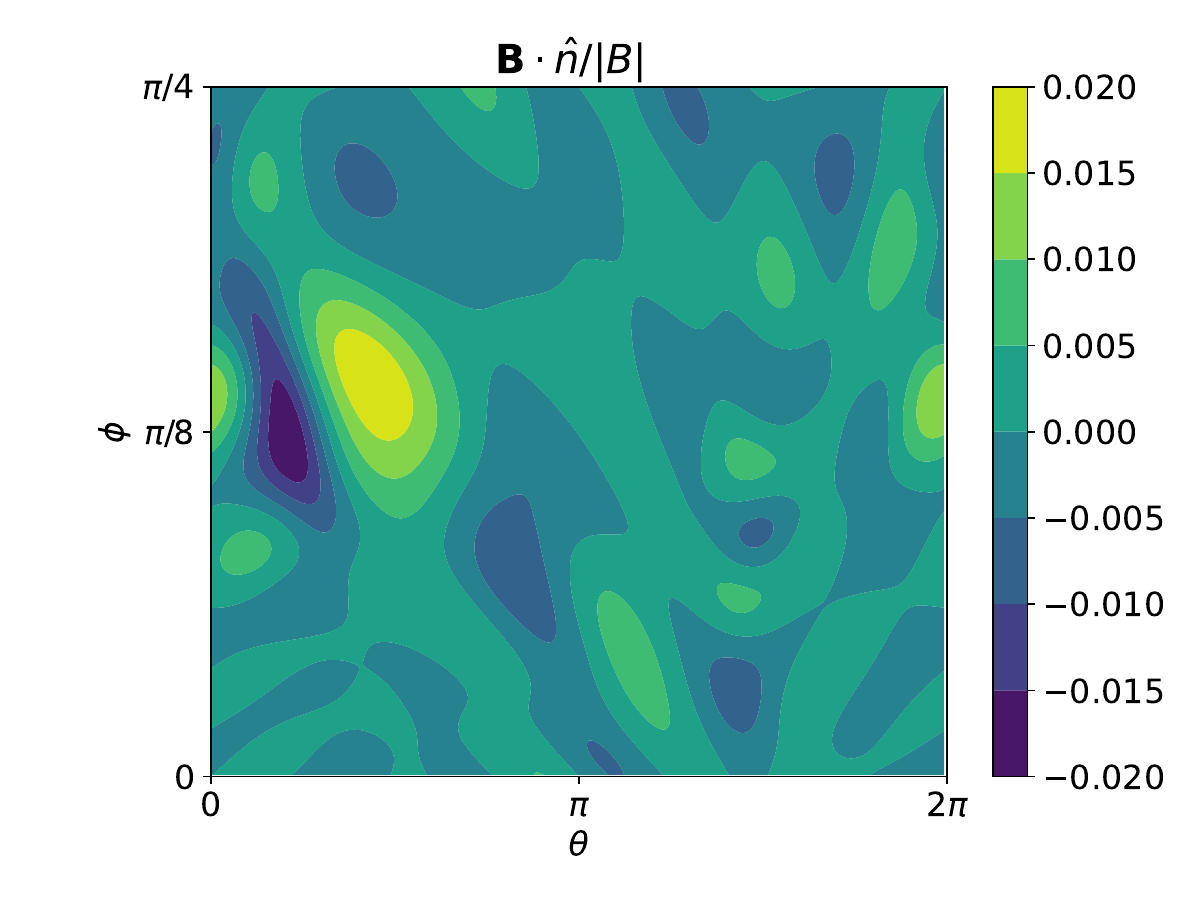}};
    \node (t2) at (0,-6) {\includegraphics[width=.8\linewidth]{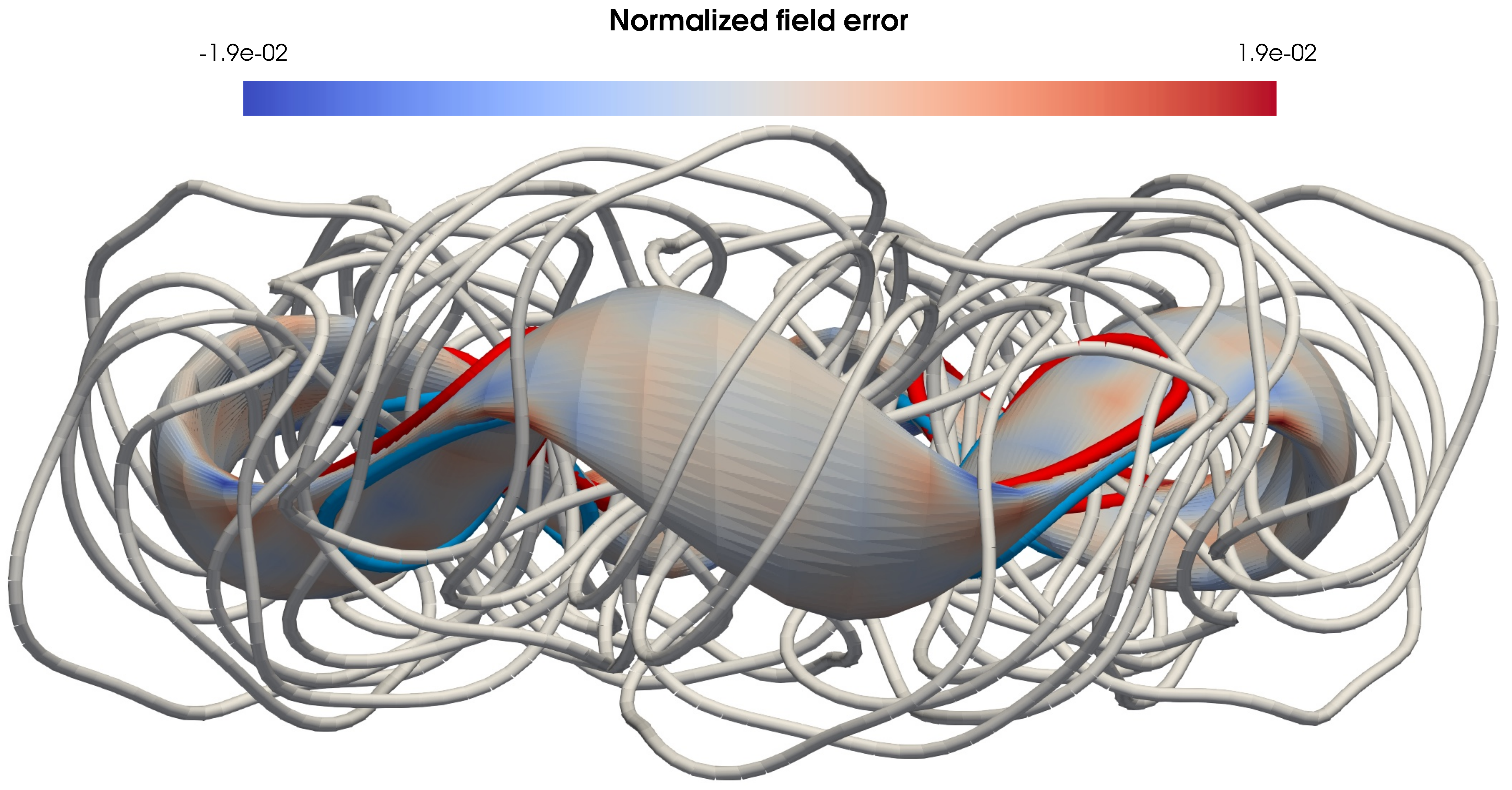}};
    \end{tikzpicture}
    \caption{Top view (top left) and side view (bottom) of the precise QH configuration with vertical access port. Coils are plotted in gray, upward-facing ports in red, downward-facing ports in blue, and the colors on the plasma boundary are the normalized field error, \textit{i.e.} $\mathbf{B}\cdot\mathbf{n}/B$. Top right: Normalized field error over one half field period.}
    \label{fig:VAQH_example}
\end{figure}

\begin{figure}
    \centering
    \begin{tikzpicture}
        \node (t1) at (-4,0) {\includegraphics[height=5cm]{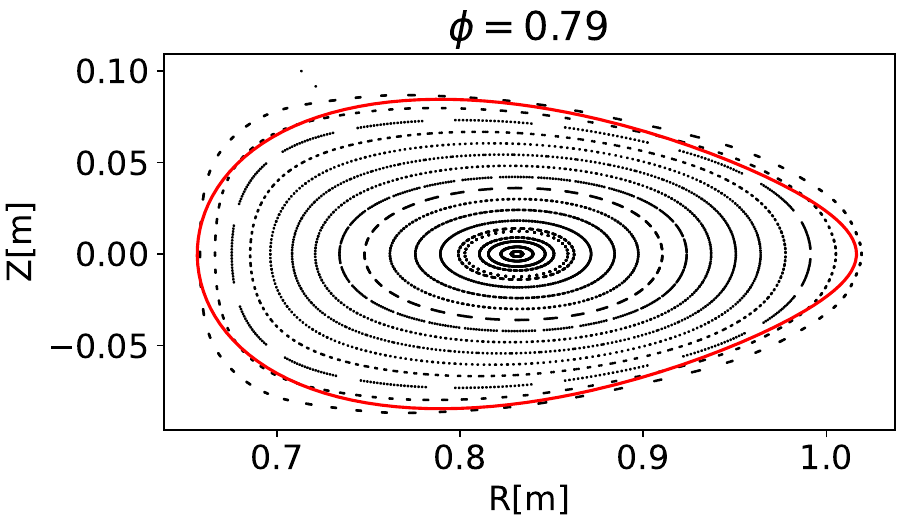}};
        \node (t2) at ( 4,0) {\includegraphics[height=7cm]{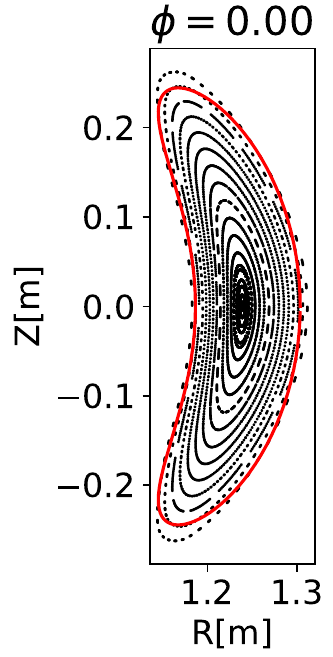}};
    \end{tikzpicture}
    \caption{Poincar\'e section (in black) at the triangular cross-section (left) and bean cross-section (right) for the precise QH configuration with vertical access port. The red curve shows the target plasma boundary.}
    \label{fig:stage_II_VAQH_poincare}
\end{figure}

We provide one last example of a stage II optimization, this time in the case of the precise QH configuration with vertical access ports. The normalized field error, as well as three-dimensional plots of the coils, the ports and the plasma boundary are shown on Figure \ref{fig:VAQH_example}, while Poincar\'e sections of the magnetic field are shown on Figure \ref{fig:stage_II_VAQH_poincare}. The obtained port has a normalized area of $A_\text{area}/aR_0 = 1.40$ for a maximum normalized field error of $\max\mathbf{B}\cdot\hat{n}/B = 2\%$. As in the case of the precise QA configuration with vertical access ports, the access port is small in comparison to radial access ports for the same configuration, and the magnetic field produces a slightly deformed plasma boundary in comparison to the target plasma boundary. Again, it is observed that obtaining large vertical access ports without compromising the field quality is difficult.

\clearpage

\bibliography{accessibility.bib}

\end{document}